# Controlled Synthesis of Organic/Inorganic van der Waals Solid for Tunable Light-matter Interactions

Lin Niu,[1,†] Xinfeng Liu,[2,†] Chunxiao Cong,[2,†] Chunyang Wu,[3,†] Di Wu,[4] Tay Rong Chang,[5] Hong Wang,[1] Qingsheng Zeng,[1] Jiadong Zhou,[1] Xingli Wang,[6] Wei Fu,[1] Peng Yu,[1] Qundong Fu,[1] Sina Najmaei,[7] Zhuhua Zhang,[8] Boris I. Yakobson,[6] Beng Kang Tay,[6] Wu Zhou,[9] Horng Tay Jeng,[5] Hsin Lin,[4] Tze Chien Sum,[2] Chuanhong Jin,[3] Haiyong He,[1*] Ting Yu,[2*] and Zheng Liu[1,6*]

[1] School of Materials Science & Engineering, Nanyang Technological University, Singapore 639798, Singapore
[2] School of Physical and Mathematical Sciences, Nanyang Technological University, Singapore 637371, Singapore
[3] State Key Laboratory of Silicon Materials and School of Materials Science and Engineering, Zhejiang University, Hangzhou 310027, P. R. China
[4] Centre for Advanced 2D Materials and Department of Physics, National University of Singapore, Singapore 117542, Singapore
[5] Department of Physics, National Tsing Hua University, Hsinchu 30013, Taiwan
[6] NOVITAS, Nanoelectronics Centre of Excellence, School of Electrical and Electronic Engineering, Nanyang Technological University, 639798, Singapore
[7] United States Army Research Laboratories, Sensors and Electron Devices Directorate, 2800 Powder Mill Road, Adelphi, MD 20783, USA
[8] Department of Materials Science and Nanoengineering, Rice University, Houston, Texas 77005, United State
[9] Materials Science and Technology Division, Oak Ridge National Lab, Oak Ridge, Tennessee 37831, USA

† L.N., X.L., C.C. and C.W. contributed equally to this work

* Corresponding authors: Z.Liu@ntu.edu.sg; YuTing@ntu.edu.sg; HYHe@ntu.edu.sg

**Van der Waals (vdW) solids, as a new type of artificial materials that consist of alternating layers bonded by weak interactions, have shed light on fascinating optoelectronic device concepts. As a result, a large variety of vdW devices have been engineered via layer-by-layer stacking of two-dimensional materials, although shadowed by the difficulties of fabrication. Alternatively, direct growth of vdW solids has proven as a scalable and swift way, highlighted by the successful synthesis of graphene/h-BN and transition metal dichalcogenides (TMDs) vertical heterostructures from controlled vapor deposition. Here, we realize high-quality organic and inorganic vdW solids, using methylammonium lead halide ($CH_3NH_3PbI_3$) as the organic part (organic perovskite) and 2D inorganic monolayers as counterparts. By stacking on various 2D monolayers, the vdW solids behave dramatically different in light emission. Our studies demonstrate that h-BN monolayer is a great complement to organic perovskite for preserving its original optical properties. As a result, organic/h-BN vdW solid arrays are patterned for red light emitting. This work paves the way for**

**designing unprecedented vdW solids with great potential for a wide spectrum of applications in optoelectronics.**

Van der Waals (vdW) solids are vertically stacked heterostructures that consist of different layered components. These materials have opened a window into a new landscape of next generation electronics and optoelectronics such as transistors,[1-5] sensors,[6-8] photodetectors,[9-14] and spin-valleytronic devices.[15-17] In light of the discoveries in inorganic two-dimension (2D) crystals, such as graphene,[5,7,18] hexagonal boron nitride (h-BN)[5,19,20] and transition metal dichalcogenides (TMDs),[21,22] one could engineer composition, thickness and stacking sequences of layered components in vdW solids for design of new material architectures and properties suited for specific applications.

Light-matter interaction at the interfaces of vdW solids is the key to build high-performance optoelectronic devices. Such interfaces have been realized in inorganic $WS_2$/graphene vdW solids.[10] In order to engineer an excellent inorganic/organic interface for the study of light-matter interaction, a suitable photoactive material needs to be identified. Over the past decade, numerous organic films such as rubrene, anthracene, pentacene, and their heterostructures have been explored active optical and electronics materials.[23-28] Recently, among plenteous photoactive materials, organic perovskites have attracted great attention due to their remarkable performance and relatively low cost as good lighter absorbers in solid-state photovoltaic devices.[29-32] Perovskites have shown highly stable excitons,[33] strong exciton absorption, sharp exciton emission even at room temperature,[34,35] high optical absorption coefficient, optimal band gap, and long electron/hole diffusion lengths[29], which grant their high efficiencies for energy conversion. The highest efficiency of organometal perovskite is ~ 20%.[36] It turns out that the performance of the organometal perovskite devices highly relies on the material quality and the surfaces. Various 2D layers could be excellent substrates for such materials and one should examine their interfacial impact, in greater details, on the optical properties of perovskite.[32] These expected complementary properties of such hybridizations suggest that marrying inorganic 2D materials with organic perovskites may provide opportunities to understand the light-matter interactions at their hetero-interfaces. Although this organic perovskite/2D heterostructure has shown promising light-based potentials in the visible regime, the difficulty of fabrication remains.

As discussed, inorganic and organic heterostructures have been developed in parallel. It is mainly because of materials themselves and fabrication processing. To date, most inorganic vdW solids such as graphene/h-BN, graphene/$MoS_2$, $MoS_2$/$WS_2$, graphene/h-BN superlattice and graphene/h-BN/$MoS_2$ stacks were engineered via layer-by-layer transfer of 2D films. Some organic films will suffer from the organic solution (e.g. acetone) used during the transfer process.[7,19,37-40] Moreover, it generally requires a few cycles of dry transfer, alignment, e-beam lithography, plasma etching and metal deposition to engineer vdW devices,[1] which may have detrimental effects on organic quality. Recently, deposition technologies such as chemical vapor deposition (CVD) have been applied to prepare vdW solids such as BN/graphene and TMD vdW heterostructures.[2,41-45] But the high reaction temperature may decompose the organic counterpart and challenges with adaptation of such growth processes remains in organic/inorganic vdW solids.

In order to address these problems, we propose a three-step method for controlled synthesis of organic/inorganic vdW solids. This protocol enables production of large-size and highly crystalline inorganic/organic vdW solid families. The inorganic monolayers could be substituted by many available 2D materials. Surprisingly, in such organic/inorganic interfaces, the sub-nm monolayer beneath the organic component can dramatically tune the optical properties of the vdW solids, by coupling with the electron-hole generated in the organic perovskites. Further studies reveal that h-BN monolayer is an excellent complement to organic perovskites for preserving their original optical properties, as also supported by theoretical calculation and the patterned organic perovskite/h-BN vdW solids as red light emitters. This work shines light on ways for designing vdW solids with great potential for a wide variety of applications in optoelectronics.

The feasibility of the strategy and method developed have been demonstrated by using the most popular carbon-based 2D material, graphene, and non-carbon 2D materials, h-BN and $MoS_2$, as the substrates for growing the organic/inorganic vdW solids. Firstly, large-size 2D monolayers are prepared as inorganic components (Figures S1-S4). Next, epitaxial growth of highly crystalline $PbI_2$ nanoplatelets on the pre-fabricated 2D substrates is performed via a physical vapor deposition (PVD) process. The $PbI_2$ crystals are then converted into perovskite by reacting with $CH_3NH_3I$ under vacuum, forming organic perovskite (methylammonium lead halide)/2D vdW solids, denoted as perovskite/2D vdW solids. The morphologies of the organic perovskite/2D vdW solids were examined by optical microscopy, scanning electron microcopy (SEM) and atomic force microscopy (AFM) (see Figures 1 and S5-S7). The $CH_3NH_3PbI_3$ nanoplatelets display different colors in the optical images, responding to different thickness. Their surface roughness is measured to be ~2.0 nm by AFM in Figures S6-7, indicating the high quality of the perovskite/2D vdW solids. Organic perovskites sitting on polycrystalline graphene, h-BN films and single crystal $MoS_2$ domains are illustrated in Figure 1g - 1i, respectively. The yellow, grey and blue balls correspond to organic molecule $CH_3NH_3^+$, Pb and I atoms, respectively. In the octahedral structure of the $PbI_2$, Pb atoms locate at the center of the halide octahedrons. During the conversion, $CH_3NH_3I$ reacts with $PbI_2$, and the $CH_3NH_3^+$ ions (orange balls in the schematic) insert into the octahedral structure of $PbI_2$ and stay at the center of eight lead halide octahedrons. XRD patterns of perovskite nanoplatelets are examined to evaluate the quality of perovskite/2D vdW solids. For converted perovskite vdW solids, peaks at 14.06°, 28.38°, 3l.74°, and 43.14° are located, assigned to (110), (220), (310), and (330), respectively, for $CH_3NH_3PbI_3$ perovskite with a tetragonal crystal structure (Figure S8-10). [46] It is worth noting that all the 2D monolayers we used in our experiments remain intact during the conversion.

The quality of as-grown organic perovskites has been further examined by selected area electron diffraction (SAED), high-resolution transmission electron microscopy (HRTEM) imaging and energy dispersive X-ray spectroscopy (EDX) (Figure 2 and Figures S11-13). HRTEM imaging indicates a highly crystalline $PbI_2$ with six-fold symmetric diffraction patterns from selected area electron diffraction (details of the TEM results of $PbI_2$ nanoplatelets can be found in Figures S11-12). Then, we investigate the crystal structure transformations with a complete conversion process from $PbI_2$ into $CH_3NH_3PbI_3$. Figure 2a shows a typical perovskite flake and the corresponding elemental mapping of carbon (C), nitrogen (N), lead (Pb), iodine (I) are presented in Figure 2b-2e, followed by an overlapped image in Figure 2f. These mapping confirm the elemental uniformity of the perovskite over the whole flake after conversion. In addition to XRD spectra, the crystal structure of the converted perovskite is evaluated by HRTEM imaging in Figure 2g and 2h. The lattice of perovskite is quite clear in Figure 2g. Inset is the corresponding fast Fourier transform (FFT) pattern from this HRTEM image along the [-120] zone axis (ZA). During the conversion process from $PbI_2$ into perovskite, with the chemical reaction of $CH_3NH_3I$ and $PbI_2$, the space group of the produces transferred from *P*3m1 for PbI2 into *I*4cm for perovskite. Figure 2h is a close-up of the area from the yellow dashed rectangle in Figure 2g. The interplanar distance of ~0.38 nm and ~0.32 nm can be determined, which are attributed to the (21-1) planes and (004) planes, respectively, with the tetragonal lattice of perovskite. The corresponding structural schematic of the perovskite are be found in Figure S13c. Apparently, along the [001] direction, there are alternative lines, similar to the HRTEM results in Figure 2g. One is consisting of I and Pb atoms while another consisting of I and $CH_3NH_3$ ions.

Confocal micro-Raman mapping and spectroscopy are commonly used powerful tools to characterize materials thanks to their unique advantages of relatively high spatial resolution (a few hundreds of nanometer), no requirement for special sample preparation, high efficiency, and non-destructiveness. Figure 3 shows the micro-Raman mapping and spectroscopy of the perovskite/2D vdW solids. Optical images of a hexagonal perovskite nanoplatelets sitting on CVD graphene, $MoS_2$, and h-BN are shown in Figs. 3a, 3d, and 3g. The thickness of the perovskite nanoplatelets on CVD graphene, $MoS_2$, and h-BN are around 300 nm, 290 nm, and 270 nm, respectively (see Supplementary Figure S9). Figures 3(b-c), 3(e-f), and 3(h-i) are the corresponding Raman mapping and spectra of the perovskite/graphene,

perovskite/MoS$_2$, and perovskite/h-BN, respectively. It can be seen from the Raman images of the sum intensity of $A_1^1$ and $A_1^2$ modes that the perovskite nanoplatelets grown on CVD graphene, MoS$_2$, and h-BN are homogenous. In the measured spectroscopy range of 10 to 160 cm$^{-1}$, the Raman spectra reveal four distinct bands at ~ 14 cm$^{-1}$, 70 cm$^{-1}$, 94 cm$^{-1}$, and 110 cm$^{-1}$, as shown in Figures 3c, 3f, and 3i, which indicate that the perovskite nanoplatelets maintain the 4H polytype of PbI$_2$ (see Figure S14).[47] The phonon vibration locates at ~ 14 cm$^{-1}$ was assigned to $E_2^3$, the shear- motion rigid-layer mode of PbI$_2$ with 4H polytype[47], while the Raman peaks at ~ 70 cm$^{-1}$, 94 cm$^{-1}$, and 110 cm$^{-1}$ were assigned to $E_2^1$, $A_1^1$, and $A_1^2$, respectively.[48] The presence of graphene, MoS$_2$ and h-BN layers underneath the perovskite are confirmed by the Raman peaks at 1590 cm$^{-1}$ (G mode of graphene), 382 and 408 cm$^{-1}$ ($E_{2g}$ and $A_{1g}$ modes of MoS$_2$), and 1369 cm$^{-1}$ ($E_{2g}$ mode of h-BN), as shown in Figure 3c, 3f, and 3i, respectively.

We also explore the UV/Vis absorption and PL excitation spectra of the perovskite/2D vdW solids on quartz. The absorption peak (760 nm) of perovskite/2D vdw solids is attributed to the direct gap transition from the valence band maximum to the conduction band minimum (Figure S15). It is found that all the absorption peaks are similar to those previously reported, where PbI$_2$ and perovskites are prepared by spinning coating and vapor deposition. These results indicate that the 2D monolayers do not change the excitation absorption of organic perovskites.[29,31] In order to examine the importance of the hetero-interface in perovskite/2D vdW solids, we characterize the samples using photoluminescence (PL) spectroscopic techniques including PL spectrum, PL mapping, and time-resolved PL measurement, as presented in Figure 4. Figures 4a-c show the PL mapping of perovskite (mapping wavelength at ~760 nm) on graphene, MoS$_2$ and h-BN, respectively. All the maps are collected under the same conditions. Their optical images are shown in Figure 3a, 3d and 3g, respectively. For comparison, the three PL mapping images are displayed using the same scale bar with the PL intensity ranging from 1 to 10$^7$. It turns out that the PL emission in perovskite is significantly modulated by the inorganic 2D components. For perovskite/graphene vdW solid, little contrast can be found. The PL emission is completely quenched (only a little intensity difference between perovskite and graphene). However, for perovskite/MoS$_2$, a sharp contrast can be observed from SiO$_2$ substrate (purple), monolayer MoS$_2$ (dark blue) and perovskite/MoS$_2$ (blue). In perovskite/h-BN, a much higher PL intensity from perovskite is recorded. The intensity is ~10$^3$ stronger than that on perovskite/graphene vdW solids. Their representative PL spectra are shown in Figure 4d. In addition, the PL intensity mapping in Figure 4c also shows distinctly different from the edge to the core region, as indicated by the white arrow, and the shape of outer layer has a consistent morphology layout as that observed in the optical microscopy image. It is ascribed to light scattering effect at edges that caused the difference of emission intensity at side and core regions. In order to understand the carrier dynamics at these distinct hetero-interfaces, time-resolved PL measurements are conducted, as shown in Figure 4e. From the single decay fitting, the lifetimes for perovskite/h-BN (green), perovskite/MoS$_2$ (pink) and perovskite/graphene (navy) vdW solids are 5.8 ns, 1.3 ns and 0.42 ns, respectively. The lifetime of the perovskite/h-BN is comparable to previous reported value of ~6.8 ns for perovskite on mica, suggesting almost no charge transfer at the interface,[49] and thus the combining rate of the electron-hope pairs remains intact on h-BN due to the very large optical band gap of h-BN (~ 6.0 eV). It is a fascinating venue to preserve the intrinsic optical properties of perovskite by using this uniform, smooth and nonconductive h-BN monolayer (see Figure 4f). In contrast, the exciton lifetime in the perovskite/MoS$_2$ interface drops remarkably, indicating that considerable charge transfer occurs at the interface. Our calculations estimate that such interface has a type II band alignment with band offset of ~0.3 eV (Figure 4g), which provides a strong driving force to separate the electron-hole pairs excited by photons. Moreover, as the electron-hole pairs in the perovskite have a small binding energy, they are particularly easy to split under a built-in electric field across the interface. Such effect dominates the carrier dynamics at perovskite/graphene interface, as evidenced by the entirely quenching behavior. In this situation, graphene has linear band dispersion with a zero band gap, which renders graphene as a collector for both electrons and holes transferred from the perovskite (Figure 4h). As such, carriers can hardly survive in the perovskites and thereby no PL peak can be measured. This efficient process for charge transfer has been demonstrated and served as electron collection layer and high performance hybrid photodetectors.[50] Therefore, tailoring the interfaces provides us an efficient way to tune and design the light-matter interactions in 2D material based heterostructures, which will benefit the fabrication of layered material based optoelectronic devices such as photodetectors, photovoltaics, light emission diodes, and on-chip lasers.

Understanding the growing mechanism is the key to improving the quality of perovskite/2D vdW solids and engineering the organic/inorganic interfaces. Ex-situ time dependent experiments are carried out to clarify this. The samples with different reaction time ($t_R$) are synthesized and characterized by optical microscopy to monitor the evolution of PbI$_2$ nanoplatelets. As shown in Figure S21, only small red dots can be observed at the early stage of the reaction ($t_R$ = 1 min). Our density functional theory (DFT) based calculations estimate that the diffusion of a PbI$_2$ nanoflake (2 nm in size) on the h-BN substrates is around 50 meV, thus being highly mobile (Figure S22a). When the reaction time is increased to 5 mins, plenty of PbI$_2$ nanoplatelets with irregular shape were formed and part of these PbI$_2$ nanoplatelets tends to coalesce with each other. The coalescence process is energetically favorable because it reduces the edge energy of PbI$_2$ nanoplatelets by decreasing their total edge length. Moreover, the rotation barrier of the nanoplatelets is also rather small, being only 45 meV for a 2 nm-diameter nanoflake on a monolayer h-BN sheet (Figure S22b). Thus, the nanoplatelets can readily adjust their crystal orientations during the coalescence process so that the misoriented interfaces (or grain boundaries) can be largely avoided. This fact agrees with our TEM characterization that the PbI$_2$ nanoplatelets can be in single crystal over tens of micrometer size. Finally, by extending the reaction time to 20 mins, the density of nuclei is decreased dramatically and hexagonal PbI$_2$ appears. On the basis of gradual morphology evolution, we propose that PbI$_2$ is formed through a van der Waals epitaxy mechanism. According to the time dependent experiments, there are two key points in the growth of perovskite/2D vdW solids: nucleation site and van der Waals epitaxy. This assumption is further supported by our perovskite/2D vdW patterns (Figure 5), where the 2D materials are patterned into arrays to separate nucleation sites (the dangling bonds within the 2D materials) from defect free surface. In a brief summary, initially, PbI$_2$ nanoparticles are nucleated around the edges and defects of 2D materials. These nanoparticles then grew along surface of 2D materials and began to merge to form highly crystalline nanoplatelets. Next, the PbI$_2$ nanoplatelets extended on 2D material surface and covered all the area. Finally PbI$_2$ will be converted into organic perovskite.

Inspired by the discussion above, based on our expertise on the controlled growth of perovskite/2D vdW and understanding of the charge transfer dynamic at the interfaces, patterned perovskite/h-BN vdW solids have been demonstrated and show excellent light emission performance as shown in Figure 5. Considering that it is energetically favorable for the PbI$_2$ to nucleate at the defects, one could control the growth of PbI$_2$ and perovskite by pre-designing the 2D films. Firstly, periodic h-BN patterns are patterned by lithography (Figure S23). Then PbI$_2$ is deposited on h-BN arrays via PVD methods. Finally, patterned perovskite/h-BN vdW solids are obtained after conversion. It is worth noting that the patterning of 2D films is compatible with traditional lithography processing. Perovskite/2D vdW solids with arbitrary geometries could be fabricated by this approach. Figure 5a presents the SEM image of perovskite/h-BN disk array in a hexagonal densely packed arrangement with patch diameter of d = 20 µm and space distance of D = 50 µm. All the h-BN patches are covered by perovskites without missing pixel, and the space between patches is very clean, indicating that our method can effectively control the growth of perovskite/2D vdW solid on targeted area. Figure 5c shows the "NTU" array derived from Figure S18b, where one can see clearly how the "NTU" array of perovskite/2D vdW solid formed. The diameter of the perovskite/h-BN hexagonal patches in the "NTU" array is around 10 µm and the gap distance between all adjacent patches is 10 µm. The gap distance between all adjacent "NTU" units is around 30 µm. Fluorescent microscopy images were taken under the excitation of irradiation (365 nm) from a high pressure mercury lamp (Figure 5b and d). Bright red luminescent patterns can be seen unambiguously without defects. The dark regions in Figures 5b and c correspond to the spaces without perovskite.

In summary, organic perovskite/2D vdW solids are prepared via a three-step process. Such vdW solids can be grown in a scalable fashion and designed into arbitrary shapes by pre-patterning the inorganic 2D monolayer using lithography. Highly crystalline organic perovskite crystals are confirmed by geometric, spectroscopic and chemical analysis. Distinct organic/inorganic interfaces are formed by using various 2D layers as inorganic components, providing an excellent scaffold for exploring and tuning the carrier dynamic at the interfaces, which is supported by our time-resolved PL experiments and DFT calculations. Based on the novel optical behaviors, complex pattern-designed perovskite/h-BN vdW solids based light emission arrays are further fabricated. This work will shed light on organic perovskite based optoelectronic devices.

**Acknowledgments**

This work was supported by the Singapore National Research Foundation under (NRF) under RF Award No. NRF-RF2013-08, the start-up funding from Nanyang Technological University (M4081137.070) and M4080514, and the Ministry of Education AcRF Tier 2 grants MOE2013-T2-1-081 and MOE2014-T2-1-044. X.L and T.C.S. also acknowledge the financial support by the Singapore NRF through the Singapore-Berkeley Research Initiative for Sustainable Energy (SinBeRISE) CREATE Programme. C.C and T. Y thanks the support of Ministry of Education - Singapore (MOE2012-T2-2-049). C.W. and C.J. thank the Center for Electron Microscopy of Zhejiang University for the access to TEM facilities, and the financial support from the National Science Foundation of China (51222202 and 51472215), the National Basic Research Program of China (2014CB932500 and 2015CB921000) and the Fundamental Research Funds for the Central Universities (2014XZZX003-07).

**Author Contributions**

Lin Niu, Xinfeng Liu, Chunxiao Cong and Chunyang Wu contributed equally to this work. Lin Niu, Xinfeng Liu and Haiyong He designed the growth and carried out part of the characterizations. Lin Niu carried out on the growth and AFM measurement. Liu Xinfeng carried out PL measurements. Cong

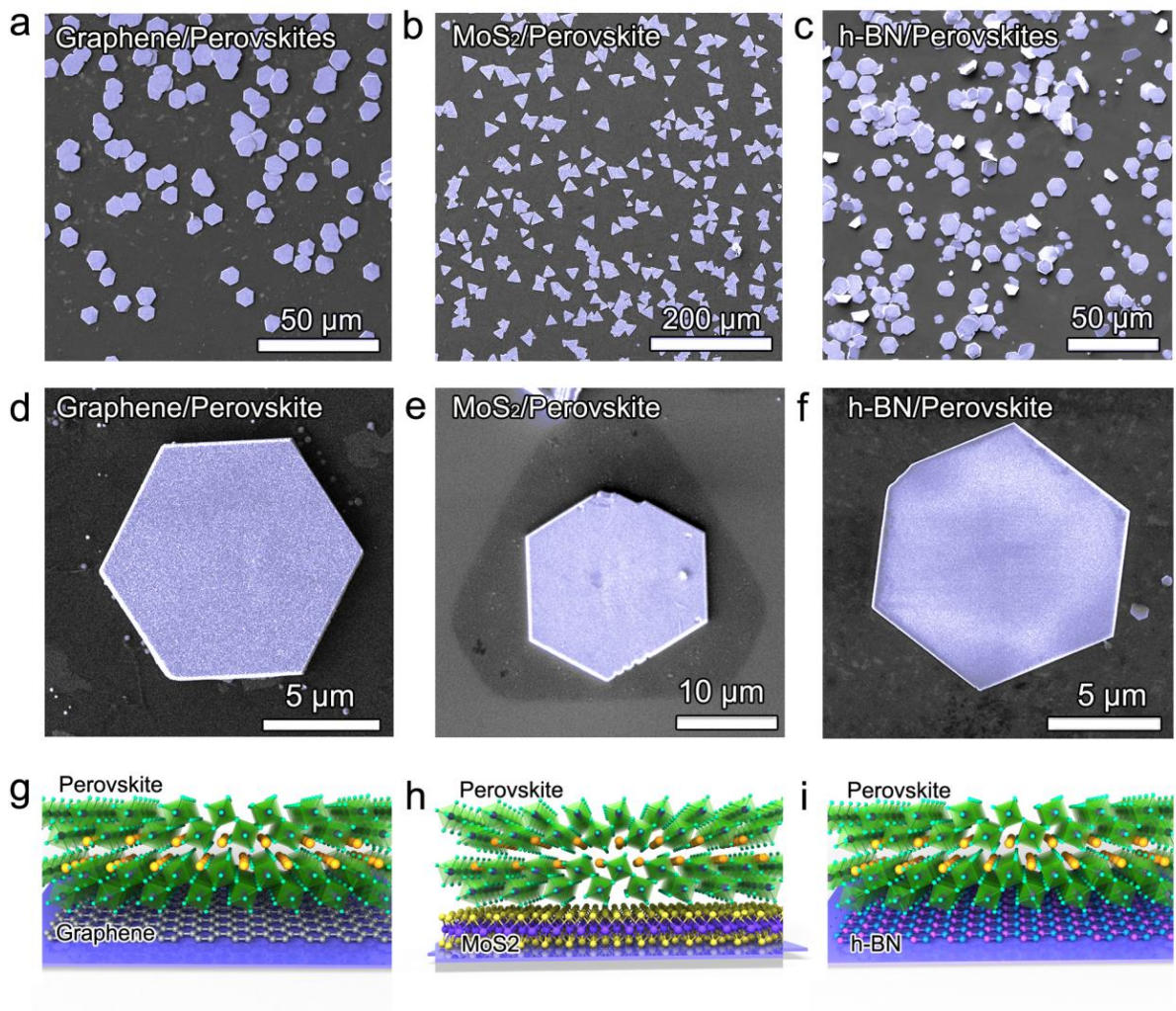

**Figure 1. Overall morphologies of the perovskite/2D vdW solids.** (a-f) False-colored SEM images of perovskite-2D vdW solids on graphene (a and d), MoS$_2$ (b and e) and h-BN (c and f), respectively. Their corresponding structure models are illustrated in g, h and i, respectively. Graphene and h-BN films cover the whole SiO$_2$/Si substrates (a, d) and (c, f) while triangular MoS$_2$ single-crystal domains are directly grown on SiO$_2$/Si substrates (b, e). Hexagonal and triangular organic perovskites are all in light blue (a – f). For the structure models (g – i), the yellow, grey and blue balls are organic molecule CH$_3$NH$_3^+$, Pb and I atoms, respectively. Pb atoms locate at the center of the halide octahedrons.

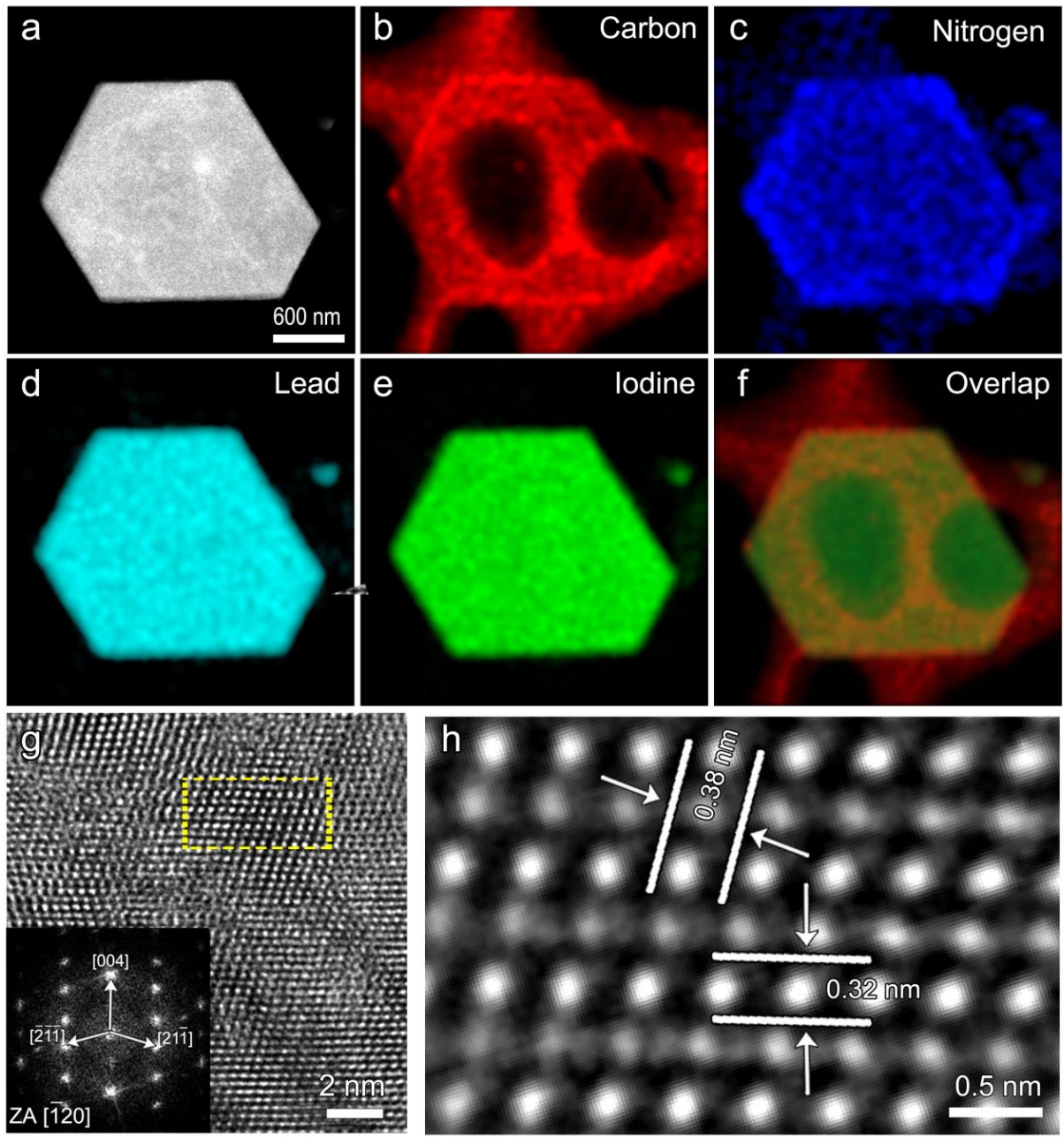

**Figure 2. Electron microscopy characterization of the perovskite/2D vdW solids.** (a) Low-resolution STEM image of a perovskite platelet. (b-f) Element mapping obtained by energy-dispersive X-ray spectroscopy show the uniformity of the perovskite nanoplatelets. (g) High resolution TEM (HRTEM) image showing the structure of the perovskite nanoplatelets. Inset is the corresponding fast Fourier transform pattern from this HRTEM image along the [-120] zone axis (ZA). (h) Filtered HRTEM image of the area highlighted in (g).

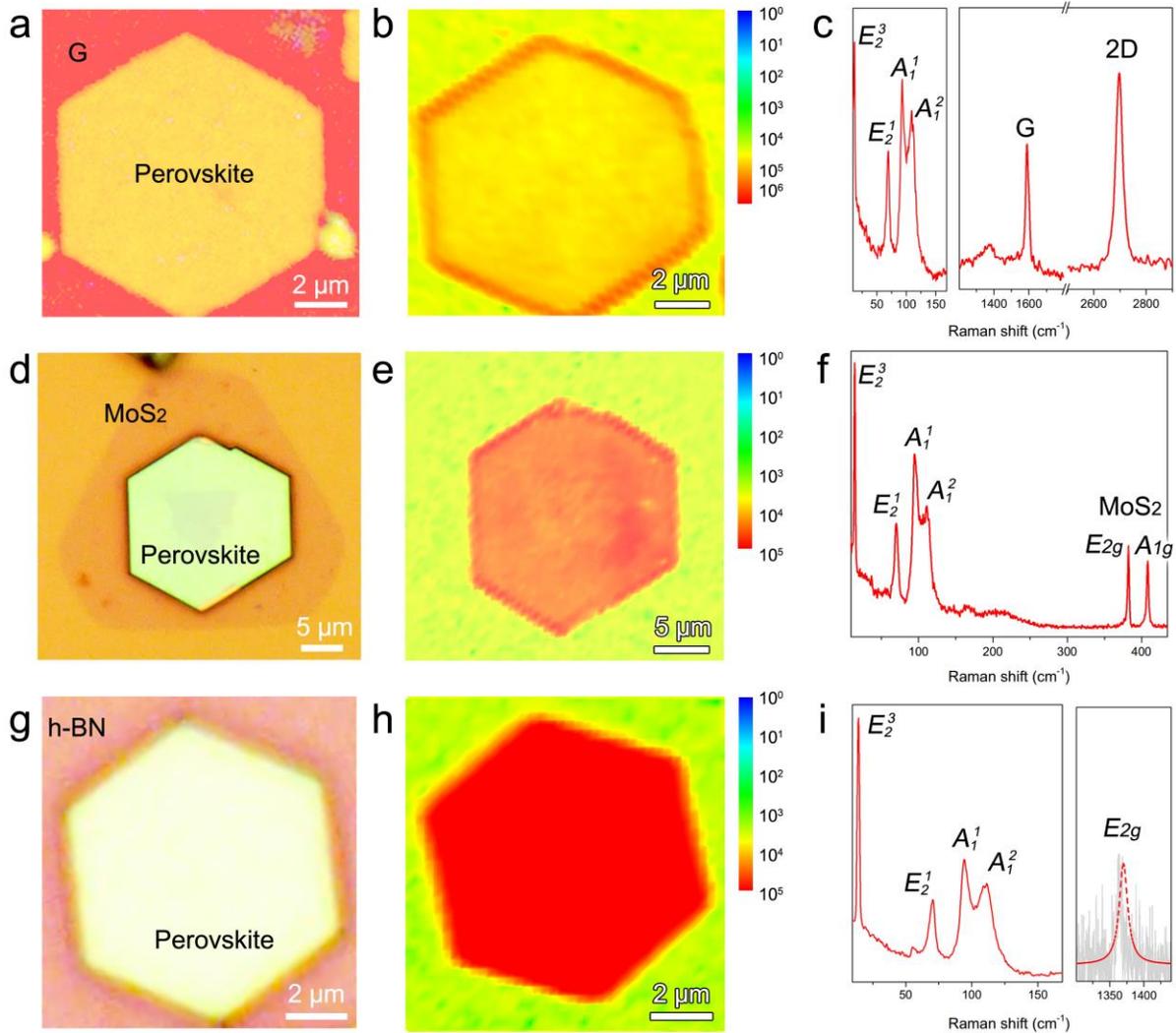

**Figure 3. Raman characterization of the perovskite/2D vdW solids.** (a, d and g) Optical images of perovskite grown on graphene, MoS$_2$ and h-BN, respectively. Scale bars are 2 μm, 5 μm and 2 μm, respectively. (b, e and h) Corresponding Raman intensity mapping of phonon modes of $A_1^1$ and $A_1^2$ of the perovskite/2D vdW solids. (c, f and i) Corresponding Raman spectra of the perovskite/2D vdW solids. The presence of graphene, MoS$_2$ and h-BN layers underneath the perovskite can be confirmed by the peaks at 1590 cm$^{-1}$ (G mode of graphene), 382 and 408 cm$^{-1}$ ($E_{2g}$ and $A_{1g}$ modes of MoS$_2$), and 1369 cm$^{-1}$ ($E_{2g}$ mode of h-BN), respectively.

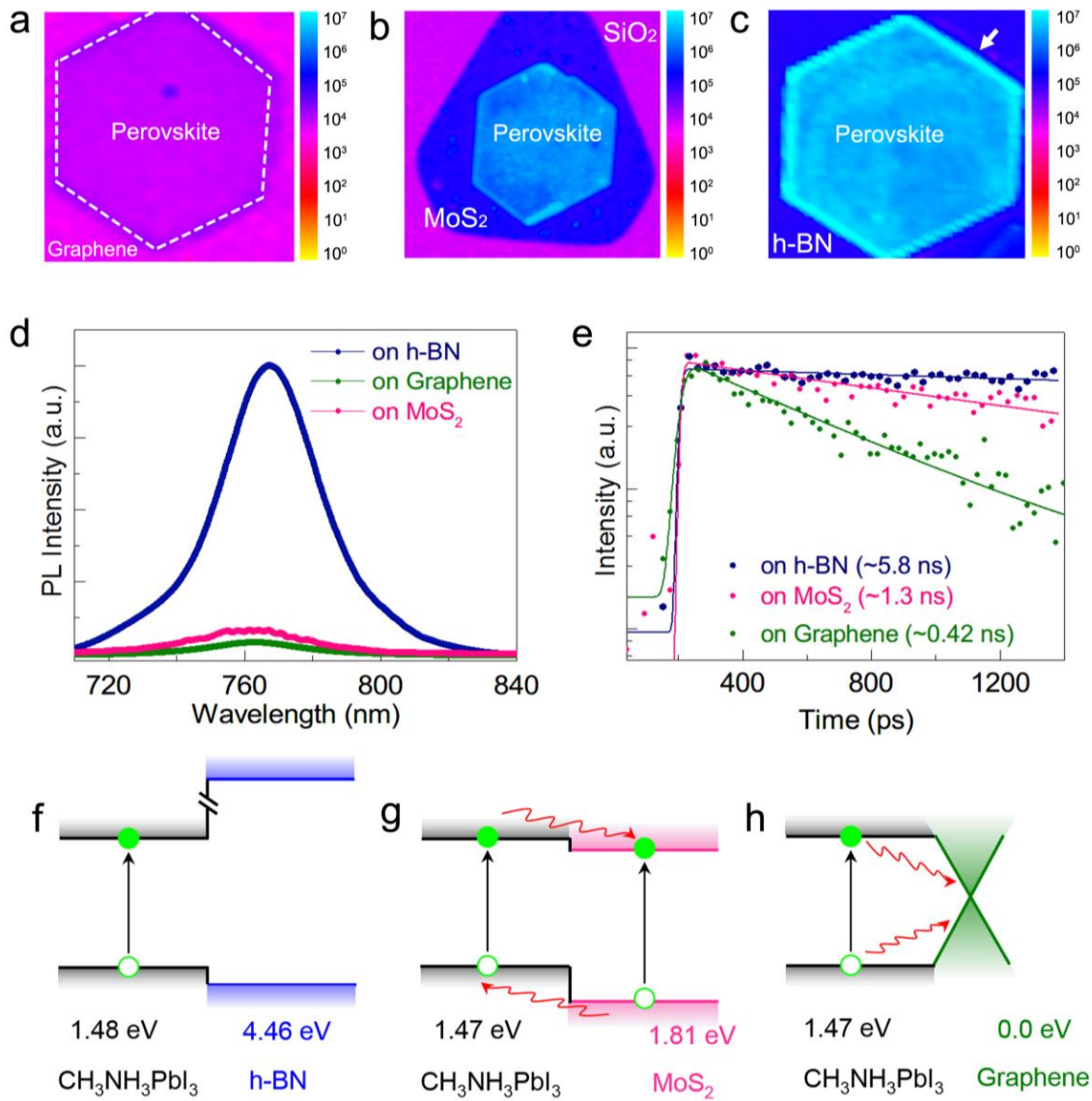

**Figure 4. Photoluminescence of the perovskite/2D vdW solids.** (a, b and c) PL intensity mapping of perovskite on graphene, MoS$_2$ and h-BN, respectively. The mapping are collected at 760 nm under the same conditions. Considerable difference in the PL intensity of the perovskites is found on graphene (purple in a, ~ 10$^4$), MoS$_2$ (blue in b, ~ 10$^6$) and h-BN (light blue in c, ~ 10$^7$), respectively. The profile of the MoS$_2$ layer is also found in dark blue in b. (d) Comparison of the PL spectra extracted from the PL mapping. (e) Time resolved PL decay transients measured at 760 ± 10 nm for perovskites on h-BN (light blue, life time of 5.8 ± 0.5 ns), MoS$_2$ (blue, life time of 1.3 ± 0.25 ns), and graphene (purple, life time of 0.42 ± 0.01 ns). The solid lines in (e) are the single-exponential fits of the PL decay transients. (f - h) Schematic band alignment and illustrative transfer of photo-excited carriers at the (f) perovskite/*h*-BN, (g) perovskite /MoS$_2$ and (h) perovskite /graphene hetero-interfaces.

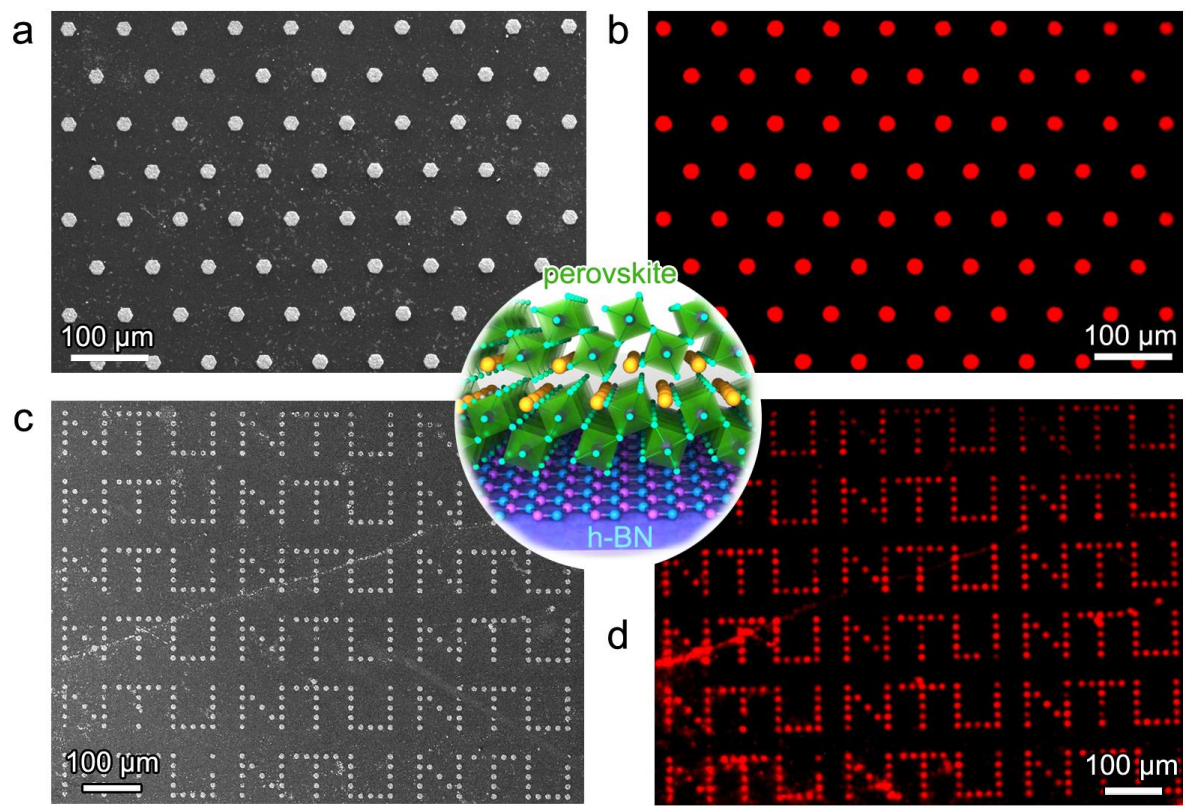

**Figure 5. Patterned perovskite/h-BN vdW solids for red light emitting array.** (a) SEM image of a perovskite/h-BN dotted array. The diameter of the dots is ~ 20 μm and the spacing is ~ 50 μm. (b) Corresponding fluorescent image showing a high intensity emission of red light. (c) SEM and (d) fluorescent image of perovskite/h-BN "NTU" arrays. The diameter of perovskite/h-BN hexagonal patches in the NTU array is ~ 10 μm and the distance between adjacent patches is ~ 10 μm. The gap distance between adjacent NTU units is around 30 μm. Center is a schematic of a perovskite/h-BN film.

# Supplementary Information

# Controlled Synthesis of Organic/Inorganic van der Waals Solid for Tunable Light-matter Interactions

Lin Niu,[1,†] Xinfeng Liu,[2,†] Chunxiao Cong,[2,†] Chunyang Wu,[3,†] Di Wu,[4] Tay-Rong Chang,[5] Hong Wang,[1] Qingsheng Zeng,[1] Jiadong Zhou,[1] Xingli Wang,[6] Wei Fu,[1] Peng Yu,[1] Qundong Fu,[1] Sina Najmaei,[7] Zhuhua Zhang,[8] Boris I. Yakobson,[6] Beng Kang TAY,[6] Wu Zhou,[9] Horng-Tay Jeng,[5] Hsin Lin,[4] Tze Chien Sum,[2] Chuanhong Jin,[3] Haiyong He,[1*] Ting Yu,[2*] and Zheng Liu[1,6*]

[1] School of Materials Science & Engineering, Nanyang Technological University, Singapore 639798, Singapore
[2] School of Physical and Mathematical Sciences, Nanyang Technological University, Singapore 637371, Singapore
[3] State Key Laboratory of Silicon Materials and School of Materials Science and Engineering, Zhejiang University, Hangzhou 310027, P. R. China
[4] Centre for Advanced 2D Materials and Department of Physics, National University of Singapore, Singapore 117542, Singapore
[5] Department of Physics, National Tsing Hua University, Hsinchu 30013, Taiwan
[6] NOVITAS, Nanoelectronics Centre of Excellence, School of Electrical and Electronic Engineering, Nanyang Technological University, 639798, Singapore
[7] United States Army Research Laboratories, Sensors and Electron Devices Directorate, 2800 Powder Mill Road, Adelphi, MD 20783, USA
[8] Department of Materials Science and Nanoengineering, Rice University, Houston, Texas 77005, United State
[9] Materials Science and Technology Division, Oak Ridge National Lab, Oak Ridge, Tennessee 37831, USA

## Part 1. Additional Experimental Details

**1. Materials Synthesis** In this section, these materials have been prepared:

-Graphene

-h-BN

-MoS$_2$

-Lead Halide Platelets-CH$_3$NH$_3$I

-CH$_3$NH$_3$PbI$_3$ Perovskite

Graphene and h-BN films were grown on 25-μm thick Cu foils (Alfa Aesar, item #46365) in a split tube furnace (Lindberg/Blue HTF55322C) with a 50 mm diameter fused quartz tube. Cu foils were pre-cleaned in dilute phosphoric acid and DI water, followed by annealing at 1000 °C for 30 min with flowing 20 sccm $H_2$ at a pressure of ~ 1.2 torr. Then 50 sccm $CH_4$ was introduced at a total pressure of ~ 1.9 torr. After 20 min growth, the furnace was cooled down naturally to room temperature under the protection of 300 sccm Ar and 5 sccm $H_2$ (total pressure: ~ 3.1 torr).

For the growth of h-BN films, Cu foils were annealed at 1025 °C for 30 min with flowing 300 sccm Ar and 50 sccm $H_2$ at a pressure of ~ 3.4 torr. After annealing, the gases were switched to 40 sccm Ar and 8 sccm $H_2$, and ammonia borane powders (placed upstream of the Cu substrates, 27 cm away from the edge of the hot furnace) were heated to 60 - 75 °C using a heating belt to support the growth of h-BN films for 30 min, the total pressure during growth is ~1.7 torr. Finally, the samples were cooled down quickly to room temperature under the protection of 300 sccm Ar and 5 sccm $H_2$ (total pressure: ~ 3.1 torr).

After growth, graphene and h-BN films were transferred to 285 nm $SiO_2$/Si substrates with the same method. Cu was removed in 0.5 M $FeCl_3$ solution under the protection of spin-coated PMMA layers, then the PMMA-supported graphene films were transferred to 285 nm $SiO_2$/Si substrate and PMMA layers were removed with acetone.

**Synthesis of CVD $MoS_2$**

The $MoS_2$ monolayers were grown by chemical vapor deposition (CVD) method under atmospheric pressure. $MoO_3$ power in a silicon boat was located in the center of a tube furnace and sulfur elemental powder in another silicon boat was located in the furnace mouth. A piece of Si wafer with 280nm $SiO_2$ top layer was suspended on $MoO_3$ boat with polished surface up.

Argon gas of 60 standard cubic centimeters per minute (sccm) was used as the carrier gas, and provided an inert atmosphere as well. The system was heated up to 750 °C in 30 min and maintained at 750 °C for 30 min. and then the system cool down naturally.

**Synthesis of Lead Halide Platelets**

$PbI_2$ powder (99.999%, Aldrich) was used as a single source and put into a quartz tube mounted on a single-zone furnace (Lindberg/Blue MTF55035C-1). The freshly CVD Graphene or $MoS_2$ or BN substrate (1 cm×3 cm) was placed in the downstream region inside the quartz tube. The quartz tube was first evacuated to a base pressure of 3 mTorr, followed by a 50 sccm flow of high purity Ar gas. The temperature and pressure inside the quartz tube were set and stabilized to desired values for each halide (390 °C and 30 Torr). In all cases, the synthesis was carried out within 30 minutes and the furnace was allowed to cool down naturally to ambient temperature.

**Synthesis of $CH_3NH_3I$**

The synthesis of $CH_3NH_3I$: 18.0 mL of methylamine (40wt% in water, Sigma) was dissolved in 100 mL of absolute ethanol (Absolute for analysis, Merck Millipores) in a 250 mL round bottom flask. 20.0 mL of hydroiodic acid (57wt% in water, Alfa aesar) was added into the solution dropwise. After that, the solution was stirred at 0 ℃ for 2 h. The raw product was obtained by removing the solvents using a rotary evaporator. A recrystallization process of the raw product, including the redissolution in 80 mL absolute ethanol and the precipitation after the addition of 300 mL diethyl ether, was carried out twice to get a much purer product. Finally, the white color powders were collected and dried at 60 ℃ for 24 h in a Vaccum Oven.

**Synthesis of Perovskites via the reaction with $CH_3NH_3I$ and $PbI_2$**

The conversions were done using a similar CVD system. CH$_3$NH$_3$I were used as a source and placed in the center of the quartz tube while CVD Graphene or MoS$_2$ or BN substrate having as-grown lead halide platelets were placed around 5–6 cm away from the center in the downstream region. The quartz tube was first evacuated to a base pressure of 3 mTorr, followed by a 50 sccm flow of high purity Ar gas. The pressure was then stabilized to 30 Torr and the temperature was elevated to 120 °C and kept there for 1 hour after which the furnace was allowed to cool down naturally to ambient temperature.

**2. Materials Characterizations**
**In this section, following technology/equipment have been accessed for materials characterization:**

- X-ray diffraction (XRD)

- SEM

- AFM

- TEM

- EDX

- Raman system

- Photoluminescence spectroscopy

- UV-Vis absorption

- Time-resolved photoluminescence spectroscopy

**XRD characterization**

X-ray diffraction pattern (2θ scans) were obtained from perovskite/2D vdW solids supported on the SiO$_2$/Si substrates using an X-ray diffractometer (XRD Shimadzu Thin Film), using Cu-Kα radiation (λ=1.54050Å) within a diffraction angle (2θ) from 5 to 60°.

**SEM**

The SEM images of perovskite flakes were obtained using JEOL JEM7600F operated at an accelerating voltage of 10 kV.

**AFM characterization**

For all the AFM experiments were performed in tapping mode under ambient conditions (Dimension ICON SPM system, Bruker, USA). Commercial silicon tips with a nominal spring constant of 40 N/m and resonant frequency of 300 kHz were used in all the experiments.

**TEM**

The $PbI_2$ samples for transmission electron microscope (TEM) were flaked off from the 2D substrates by using Toluene (99.85%, Acros Organics) and then transfer onto the TEM grids (Quantifoil Mo grids).The $CH_3NH_3PbI_3$ perovskite samples for TEM measurement were converted from the $PbI_2$ onto the TEM grids, with the similar method we introduced in the materials synthesis part above.

The high resolution transmission electron microscopy (HRTEM) and the selected area electron diffraction (SAED) pattern were done with a FEI Tecnai F20 operated with an acceleration voltage of 200 kV. The chemical composition of lead iodide and $CH_3NH_3PbI_3$ perovskite was determined by means of energy dispersive X-ray spectroscopy (EDX attached to FEI Tecnai F20).

**Raman and PL characterization**

A WITec alpha300 RAS Raman system with a piezocrystal controlled scanning stage, an objective lens of ×100 magnification (numerical aperture, NA = 0.95), and an Electron Multiplying CCD was used for recording photoluminescence (PL) and Raman spectra/images.

For PL spectra/images, a 600 lines/mm grating was used. For Raman spectra/images, a low-wavenumber coupler and a 2400 lines/mm grating were used for observing low-frequency Raman modes and achieving a good spectral resolution. All the PL and Raman spectra/images were recorded under an excitation laser of 532 nm ($E_{laser}$ = 2.33 eV). To avoid the laser-induced heating, laser power was kept below 0.1 mW. The laser spot is of ~500 nm in diameter.

**UV-Vis absorption characterization**

UV-Vis absorption spectra of perovskite/2D vdW solids prepared on quartz were recorded on SHIMADZU UV-3101PC UV–vis-NIR scanning spectrophotometer.

**Time-resolved photoluminescence spectroscopy**

The excitation pulse (400 nm) was generated by frequency doubling the 800 nm output (with a BBO crystal) from the Coherent Oscillator Mira 900F (120 fs, 76 MHz, 800 nm). The pump laser source is introduced into a microscope (Nikon LV100) and focused onto samples via a 20× objective (Nikon, numerical aperture: 0.4). The PL emission signal was collected in a standard backscattering geometry and dispersed by a 0.25 m DK240 spectrometer with 150 g/mm grating. The emission signal was time-resolved using an Optoscope Streak Camera system that has an ultimate temporal resolution of ~10 ps.

**FL characterization**

The fluorescence images were obtained by an Olympus fluorescence microscope. A Mercury lamp was used as the excitation light source. All FL images are as-taken without any artificial image processing.

**3. h-BN patterns Fabrication**

As-transferred h-BN film can be used to prepare various patterns for inducing the growth of PdI$_2$ patterns. The fabrication process consists of the following steps. (1) The h-BN film was covered by a photoresist layer spin-coated (AZ5214, 4000 r/min) on top of the h-BN surface. (2) Standard photolithography is performed to pattern the photoresist layer as a mask. (3) Argon-based plasma etching (power is 50 W, pressure is 200 mTorr and time is 30 s) is performed to transfer the photoresist mask pattern onto underlying h-BN. (4) Photoresist mask is completely removed in acetone, and a h-BN pattern is created, such as the hexagon and "NTU" patterns shown in Fig. S21.

## 4. DFT Calculations

Our results are based on first principle calculations within density functional theory (DFT) as implemented in the VASP[1] (Vienna *ab-initio* simulation package) code. The ion-electron interaction was modeled by projector augmented plane wave (PAW)[2] and the electron exchange correlation was treated by Perdew-Burke-Ernzerhof (PBE) parameterization of the generalized gradient approximation (GGA)[3]. The plane wave basis set with kinetic energy cutoff of 500 eV was used. The Brillouin zone was sampled by a 8×8×1 Γ-centered k-point mesh. A vacuum layer of 15 Å was adopted to avoid interactions between the neighbor surfaces. All structures were fully relaxed until the Hellmann-Feynman force on each atom was smaller than 0.01 eV/Å.

Potential energy calculations: The energies are calculated by first-principles calculations as implemented in VASP code as well. We employed ultrasoft pseudo-potentials for the core region and spin-unpolarized density functional theory based on local density approximation, which could give a reasonable interlayer distance between the PbI$_2$ flake and *h*-BN sheet. A kinetic energy cutoff of 400 eV is chosen for the plane-wave expansion.

**Part 2. Additional Figures and Discussions**

**S1-15. Synthesis and additional characterizations of perovskite/2D vdW solids**

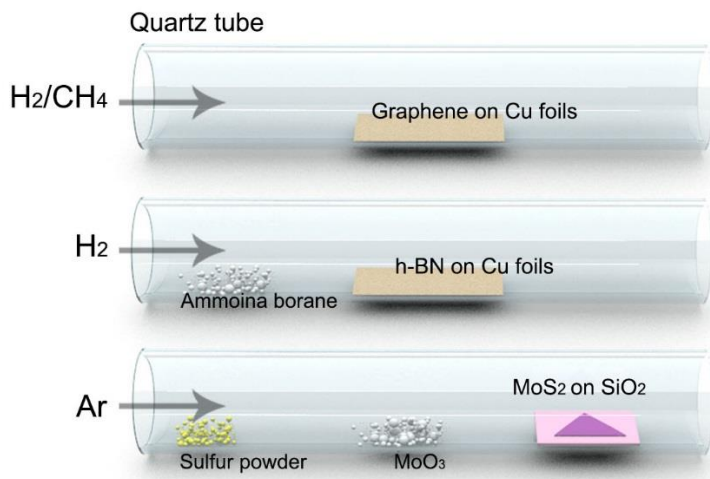

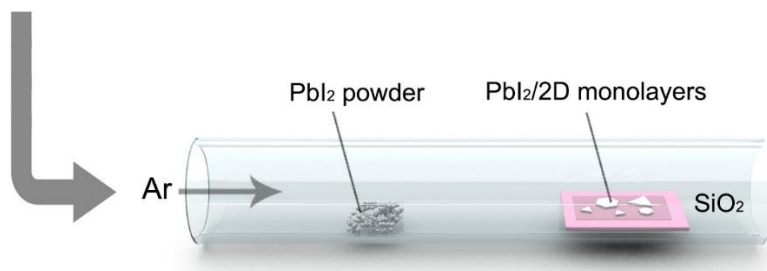

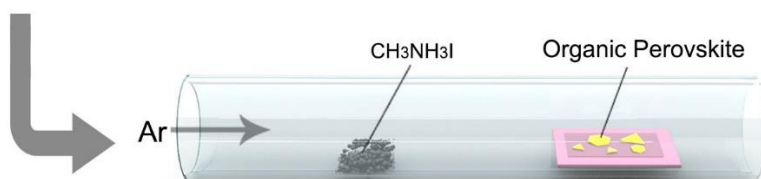

**Figure S1. Three-step method for synthesis of perovskite/2D vdW solids.** On step 1, large-size 2D monolayers including graphene, MoS$_2$ and h-BN are developed, separately, as inorganic components. Next, on step 2, epitaxial growth of highly crystalline PbI$_2$ nanoplatelets on these pre-fabricated 2D substrates are performed via a physical vapor deposition (PVD) process. The PbI$_2$ crystals are then converted into perovskite on step 3, whereby reacting with CH$_3$NH$_3$I under vacuum.

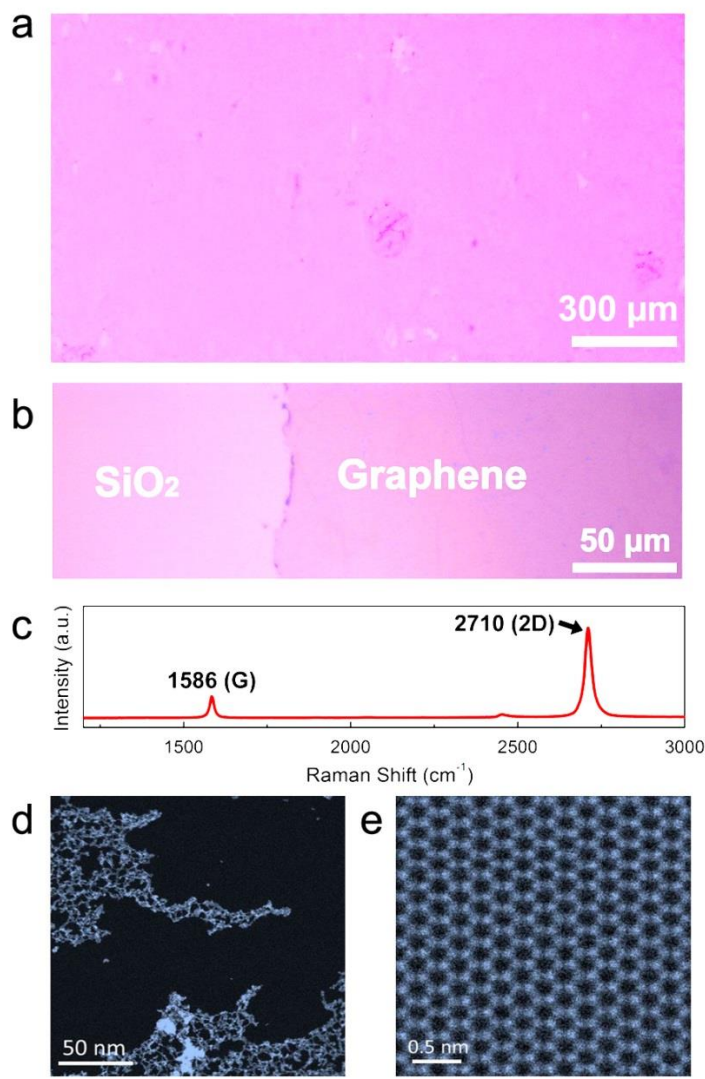

**Figure S2. Characterizations of monolayer graphene grown by CVD.** (a) CVD-grown graphene was transferred on SiO$_2$/Si wafer with 285 nm thick SiO$_2$. (b) Obvious boundary of the area of CVD Graphene on SiO$_2$/Si wafer. CVD graphene monolayer has a little color contrast comparing with the SiO$_2$/Si substrate. (c) Raman characterization of CVD graphene. The Raman peaks at ~ 1586 cm$^{-1}$ and ~ 2710 cm$^{-1}$ were assigned to the G peak and 2D peak from graphene. The intensity ratio of 2D/G is ~ 4, indicating a monolayer graphene. (d) Low-magnification STEM Z-contrast image showing the presence of large area clean monolayer graphene (dark regions in the image). The bright features are surface contamination induced mostly during TEM sample preparation. (e) High resolution STEM Z-contrast image showing the hexagonal atomic arrangement in monolayer graphene.

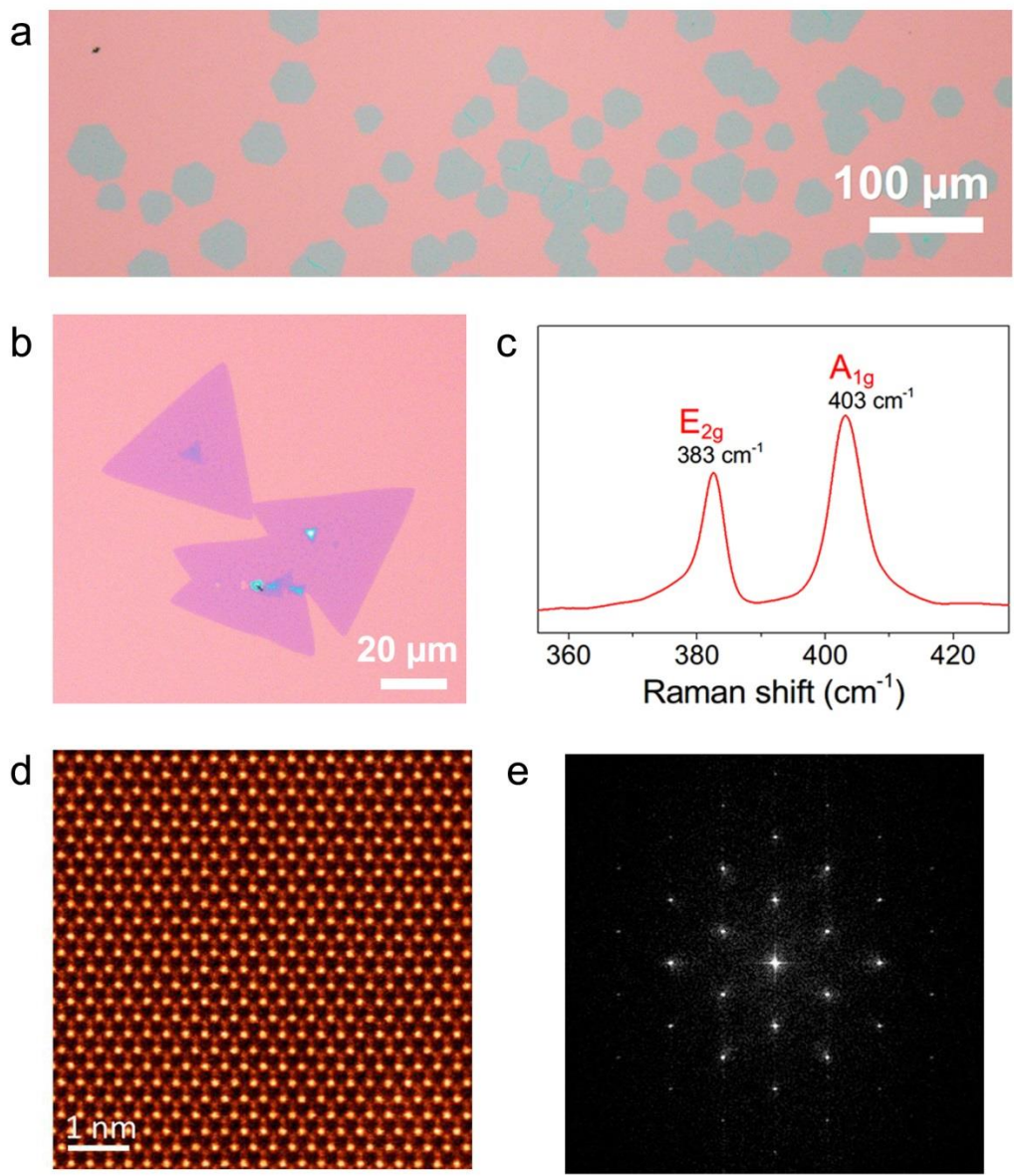

**Figure S3. Characterizations of monolayer MoS$_2$ grown by CVD.** (a-b) Optical images of MoS$_2$ at different magnifications, showing the high yield of this direct-growth method via CVD on SiO2/Si wafer with 285 nm thick SiO$_2$. (c) Raman characterization of CVD MoS$_2$. The Raman peaks at ~ 383 cm$^{-1}$ and ~ 403 cm$^{-1}$ were assigned to the E$_{2g}$ peak and A$_{1g}$ peak from MoS$_2$. The distance between two modes is ~ 20 cm$^{-1}$, suggesting a monolayered MoS$_2$. (d-e) High resolution STEM Z-contrast image and its FFT pattern showing the high crystallinity of the monolayer MoS$_2$.

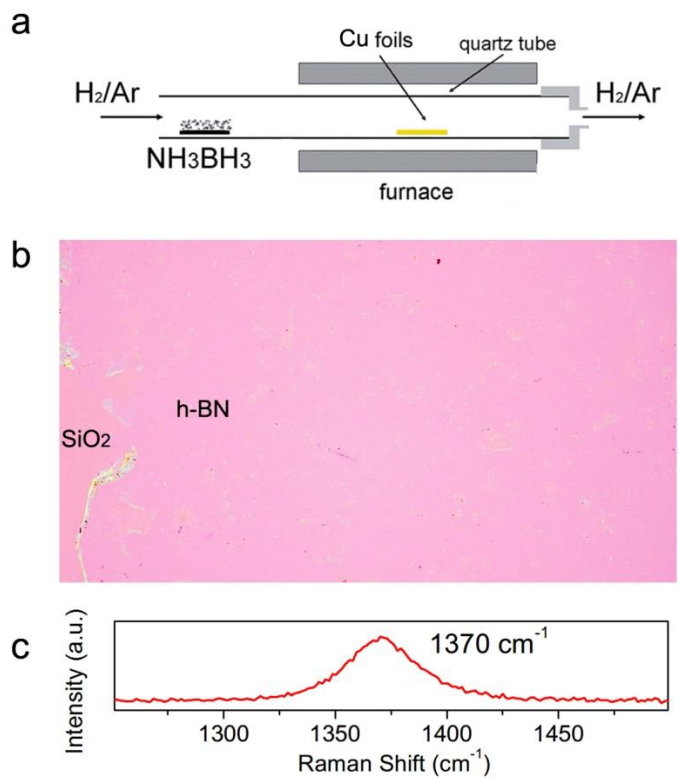

**Figure S4. Characterizations of monolayer h-BN grown by CVD.** (a) Schematic of the synthesis process of epitaxial growth of h-BN on the Cu substrates. (b) CVD-grown h-BN was transferred on $SiO_2$/Si wafer with 285 nm thick $SiO_2$. The good coverage of h-BN on SiO2 (> 90%) enables the fabrication of h-BN patterns (Figure 6 in manuscript) for light emission applications of organic perovskite/h-BN arrays. (c) Typical Raman spectrum of CVD h-BN. The Raman peak at ~ 1370 cm$^{-1}$ was assigned to the $E_{2g}$ peak from h-BN.

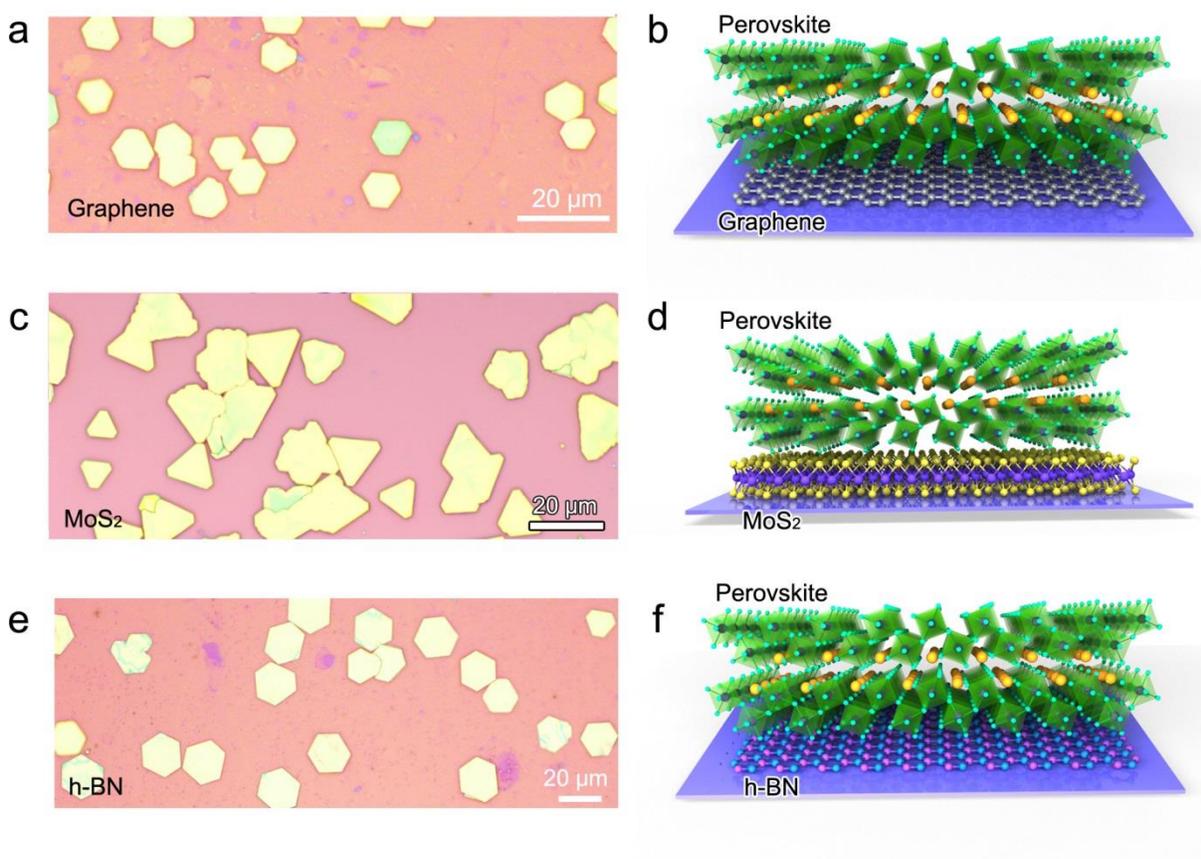

**Figure S5. Optical images and schematic of perovskite/2D vdW solids.** The morphologies of organic perovskite/2D vdW solids were examined by optical microscopy, on graphene (a), MoS2 (c) and h-BN (e), respectively. Organic perovskites are in yellow, sitting on polycrystalline graphene and h-BN films and single crystal MoS$_2$ domains, as illustrated in b, d and f, respectively. The yellow, grey and blue balls correspond to organic molecule $CH_3NH_3^+$, Pb and I atoms, respectively. In the octahedral structure of the PbI$_2$, Pb atoms locate at the center of the halide octahedrons. During the conversion, CH$_3$NH$_3$I reacts with PbI$_2$, and the CH$_3$NH$_3^+$ ions (orange ball in the schematic) insert into the octahedral structure of PbI$_2$ and stay at the center of eight lead halide octahedrons.

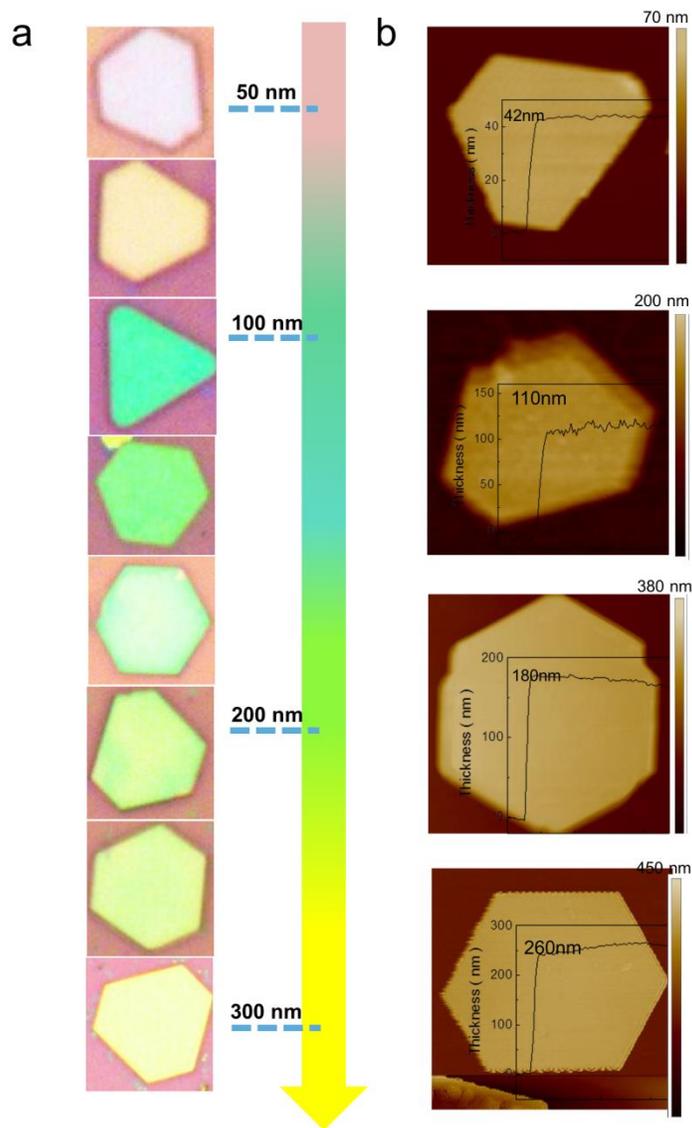

**Figure S6. Optical microscopy and AFM images of perovskite nanoplatelets with different thickness.** Column **a** (left) is the optical microscopy images. The color and brightness of perovskite nanoplatelets vary with thickness, because of the change of absorbance and optical pathway of reflected light. When the thickness of perovskite nanoplatelets changes from dozens of nanometers to a few hundred nanometers, the color varies from pink to light yellow. Column **b** (right) shows the corresponding AFM images with height and roughness profiles. Comparing the optical and AFM images, we could confirm that the color and brightness of perovskites nanoplatelets indeed have relationship with height. The roughness profiles are also shown in column b, which indicate that the surface of perovskite nanoplatelets is ultra-smooth, coinciding with the results of SEM characterization.

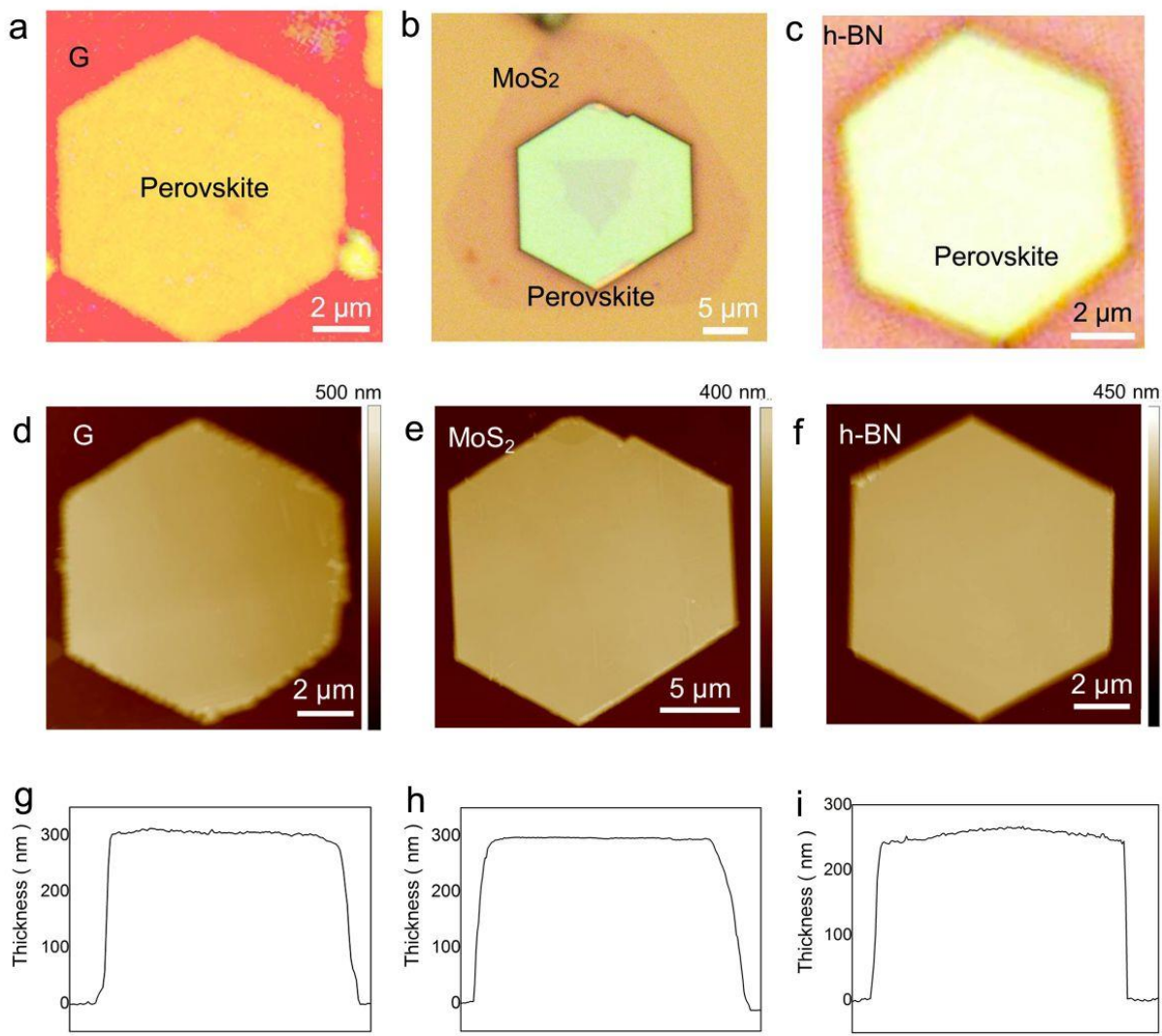

**Figure S7. Optical microscopy images and corresponding AFM data of perovskite/2D vdW solid.** (a) Optical microscopy image of perovskite/2D vdW solid grown on graphene. The red area is graphene and the yellow hexagon area is perovskite. (d) and (g) Corresponding AFM image and height profiles of (a), which demonstrate that the height of perovskite on graphene is about 300 nm and the surface of perovskite is ultra-smooth. (b) Optical microscopy image of perovskite/2D vdW solid grown on $MoS_2$. The hexagon-shape $MoS_2$, showing poor optical contrast, can be clearly seen under perovskite. (e) and (h) Corresponding AFM image and height profile of perovskite shown in (b). The height of perovskite on $Mos_2$ surface is about 290 nm. Even though there is a small triangle-shape $MoS_2$ at the center of perovskite, the surface roughness does not increase. (c) Optical microscopy image of perovskite/2D vdW solid grown on h-BN, in which the pink area is the h-BN the bright hexagon is perovskite. The corresponding AFM image and height profile are demonstrated in (f) and (i) respectively. The perovskite nanoplate on h-BN has very flat surface with a uniform thickness (270 nm) across the lateral dimension.

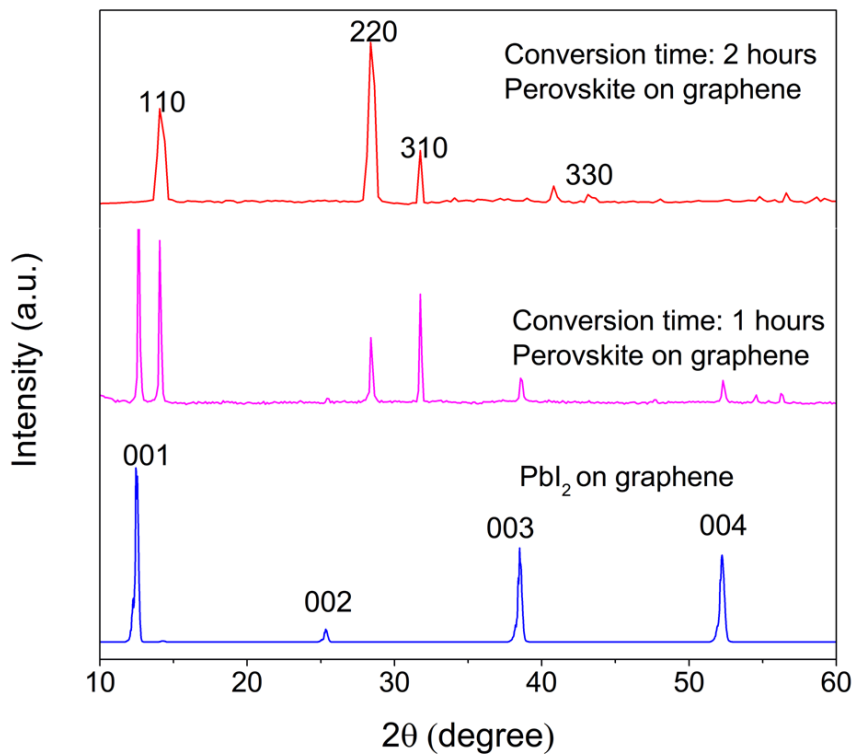

**Figure S8. Time evolution X-ray diffraction patterns of perovskite/Graphene vdW solid.** It can be seen that a set of strong peaks at 12.54, 25.34, 38.50, and 52.24, assigned to 001, 002, 003, and 004 of the $PbI_2$ crystal growing on CVD graphene, indicating high level of phase purity of hexagon crystal structure of $PbI_2$ with a highly oriented growth direction along the c-axis.[4] Further, XRD patterns are detailed accompanied by perovskite nanoplatelet evolution with different conversion time. During the perovskite conversion process (1h), corresponding peaks of both $PbI_2$ and $CH_3NH_3PbI_3$ could be obtained in XRD pattern at the $PbI_2$ and perovskite coexist stage. Peaks for $PbI_2$ nanoplatelets were still similar like the pattern of the initial stage. The perovskite have been converted in vdW solids, owning the characteristic peaks at 14.06°, 28.38°, 3l.74°, and 43.14°, assigned to (110), (220), (310), and (330) for $CH_3NH_3PbI_3$ perovskite with a tetragonal crystal structure.[5] With the advance of time, the $PbI_2$ peaks totally disappears at the conversion complete stage (2h in this work), and the pure perovskite/2D vdW solids were obtained with high crystallinity.

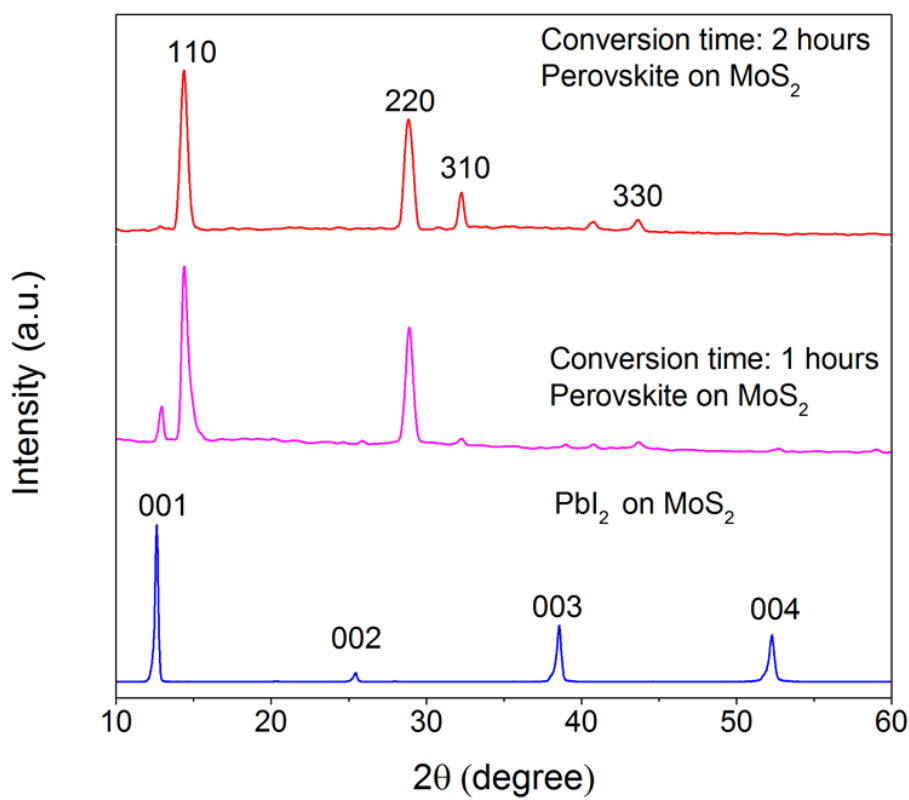

**Figure S9. Time evolution X-ray diffraction patterns of perovskite/MoS$_2$ vdW solid.** It can be seen that a set of strong peaks at 12.62, 25.48, 38.58, and 52.28, assigned to 001, 002, 003, and 004 of the PbI$_2$ crystal growing on CVD MoS$_2$, indicating high level of phase purity of hexagon crystal structure of PbI$_2$ with a highly oriented growth direction along the c-axis.[4] Further, XRD patterns are detailed accompanied by perovskite nanoplatelet evolution with different conversion time. During the perovskite conversion process (1h), corresponding peaks of both PbI$_2$ and CH$_3$NH$_3$PbI$_3$ could be obtained in XRD pattern at the PbI$_2$ and perovskite coexist stage. Peaks for PbI$_2$ nanoplatelets were still similar like the pattern of the initial stage. The perovskite have been converted in vdW solids, owning the characteristic peaks at 14.66°, 28.90°, 32.26°, and 43.65°, assigned to (110), (220), (310), and (330) for CH$_3$NH$_3$PbI$_3$ perovskite with a tetragonal crystal structure.[5] With the advance of time, the PbI$_2$ peaks totally disappears at the conversion complete stage (2h in this work), and the pure perovskite/2D vdW solids were obtained with high crystallinity.

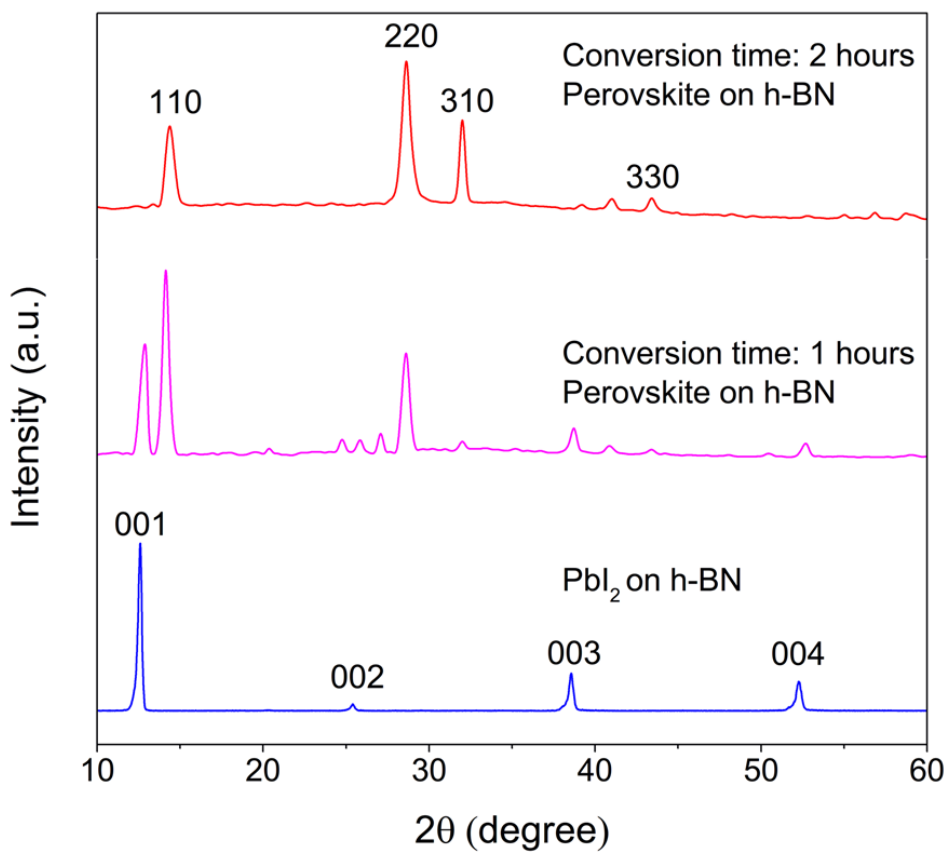

**Figure S10. Time evolution X-ray diffraction patterns of perovskite/h-BN vdW solid.** It can be seen that a set of strong peaks at 12.60, 25.40, 38.56, and 52.26, assigned to 001, 002, 003, and 004 of the $PbI_2$ crystal growing on CVD h-BN, indicating high level of phase purity of hexagon crystal structure of $PbI_2$ with a highly oriented growth direction along the c-axis.[4] Further, XRD patterns are detailed accompanied by perovskite nanoplatelet evolution with different conversion time. During the perovskite conversion process (1h), corresponding peaks of both $PbI_2$ and $CH_3NH_3PbI_3$ could be obtained in XRD pattern at the $PbI_2$ and perovskite coexist stage. Peaks for $PbI_2$ nanoplatelets were still similar like the pattern of the initial stage. The perovskite have been converted in vdW solids, owning the characteristic peaks at 14.14°, 28.63°, 32.00°, and 43.40°, assigned to (110), (220), (310), and (330) for $CH_3NH_3PbI_3$ perovskite with a tetragonal crystal structure.[5] With the advance of time, the $PbI_2$ peaks totally disappears at the conversion complete stage (2h in this work), and the pure perovskite/2D vdW solids were obtained with high crystallinity.

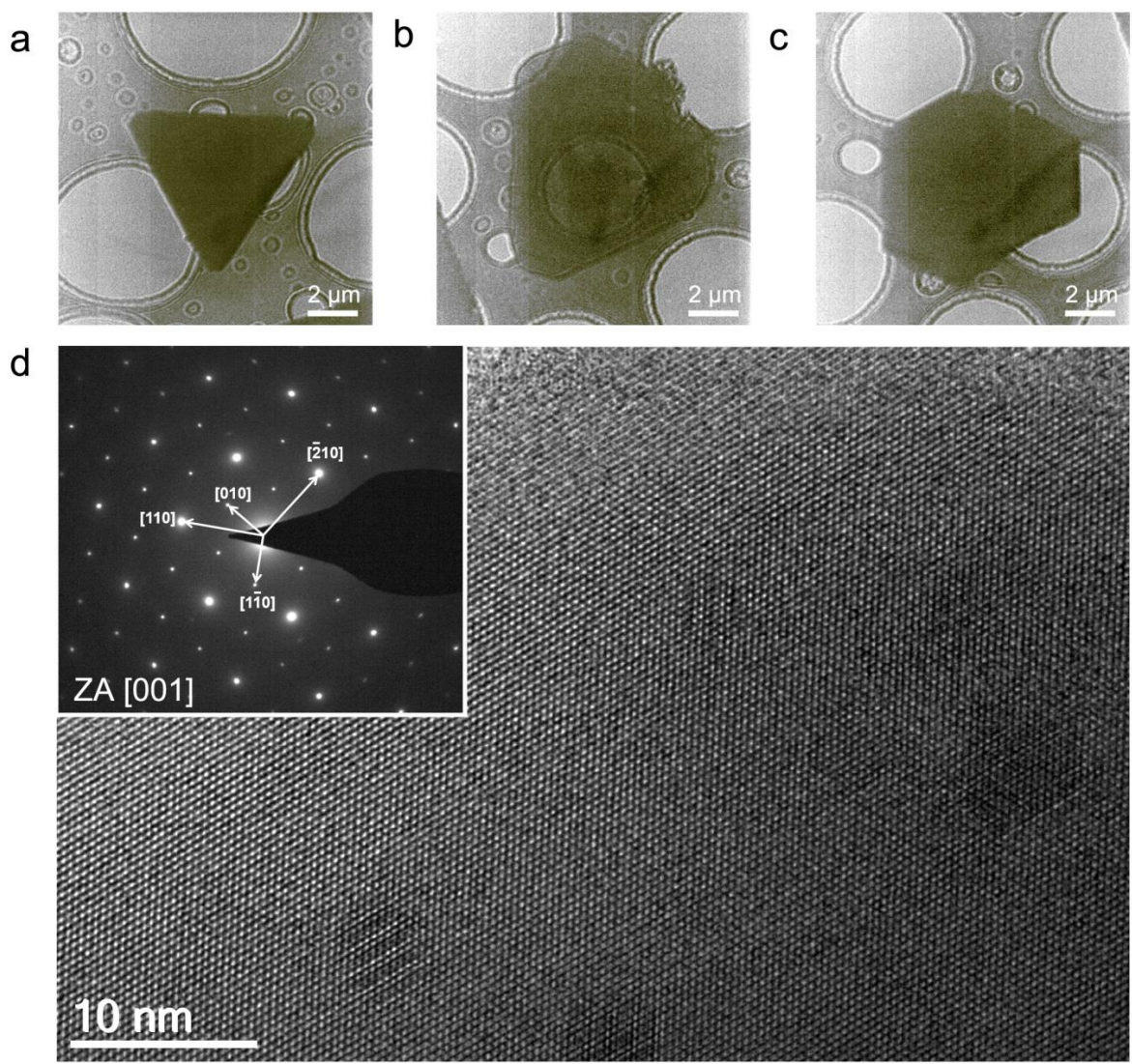

**Figure S11. Hexagonal PbI2 crystal nanoplatelets and corresponding TEM images.** (a-c) Low-resolution TEM images of $PbI_2$ nanoplatelets. (d) High resolution TEM (HRTEM) image showing the hexagonal structure of the $PbI_2$ nanoplatelets. Top inset is the selected-area electron diffraction pattern along the [001] zone axis (ZA).

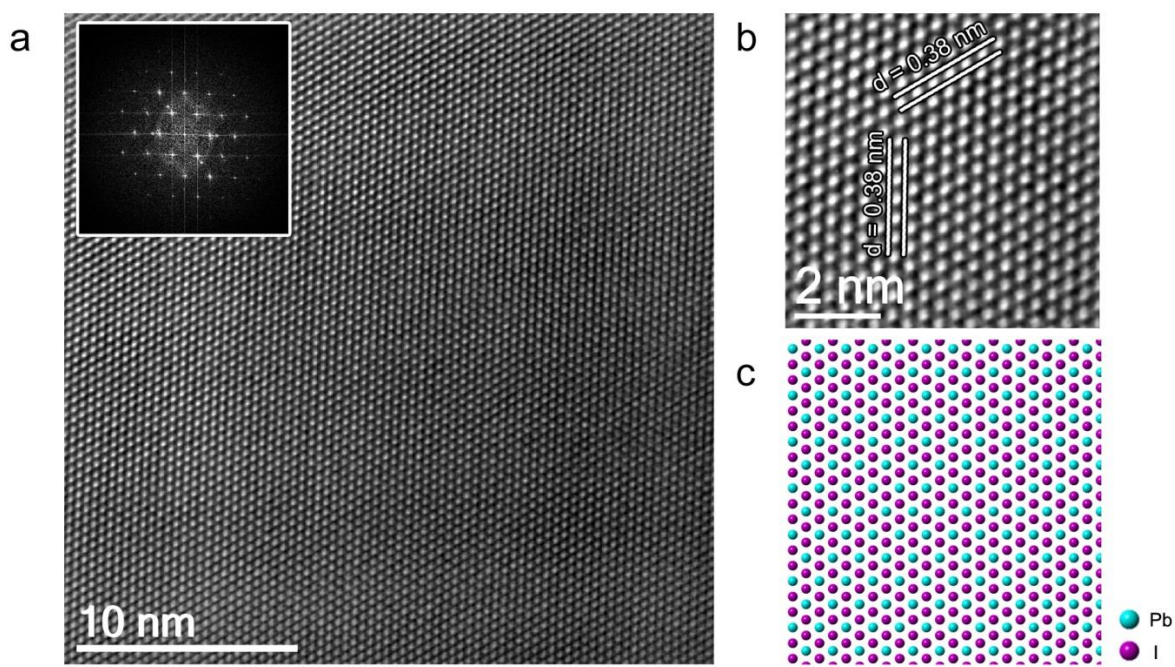

**Figure S12. High resolution TEM measurement on hexagonal PbI$_2$ crystal nanoplatelets.** (a) High resolution TEM (HRTEM) image showing the hexagonal structure of the PbI$_2$ nanoplatelets. Top inset is the corresponding fast Fourier transform pattern from this HRTEM image along ZA [001]. (b) HRTEM image of a selected-area from (a), with HRTEM filter process. (c) A structure scheme corresponding with the HRTEM image in (b). HRTEM imaging indicates a highly crystalline PbI$_2$ with six-fold symmetric diffraction patterns from selected area electron diffraction (SAED) (Figure S10d), shown in the top inset can be indexed to the high crystallinity with the zone axe (ZA) [001]. After the HRTEM filter process, a representative HRTEM image (Figure S11b) shows an interplanar distance of ~0.38 nm, which can be attributed to the (100) family planes (d100=0.39nm) with the hexagonal lattice of PbI$_2$. In the HRTEM images, the Pb atom is encircled by iodine atoms, which is in good agreement with our HRTEM image simulations as shown in Figure S11c.

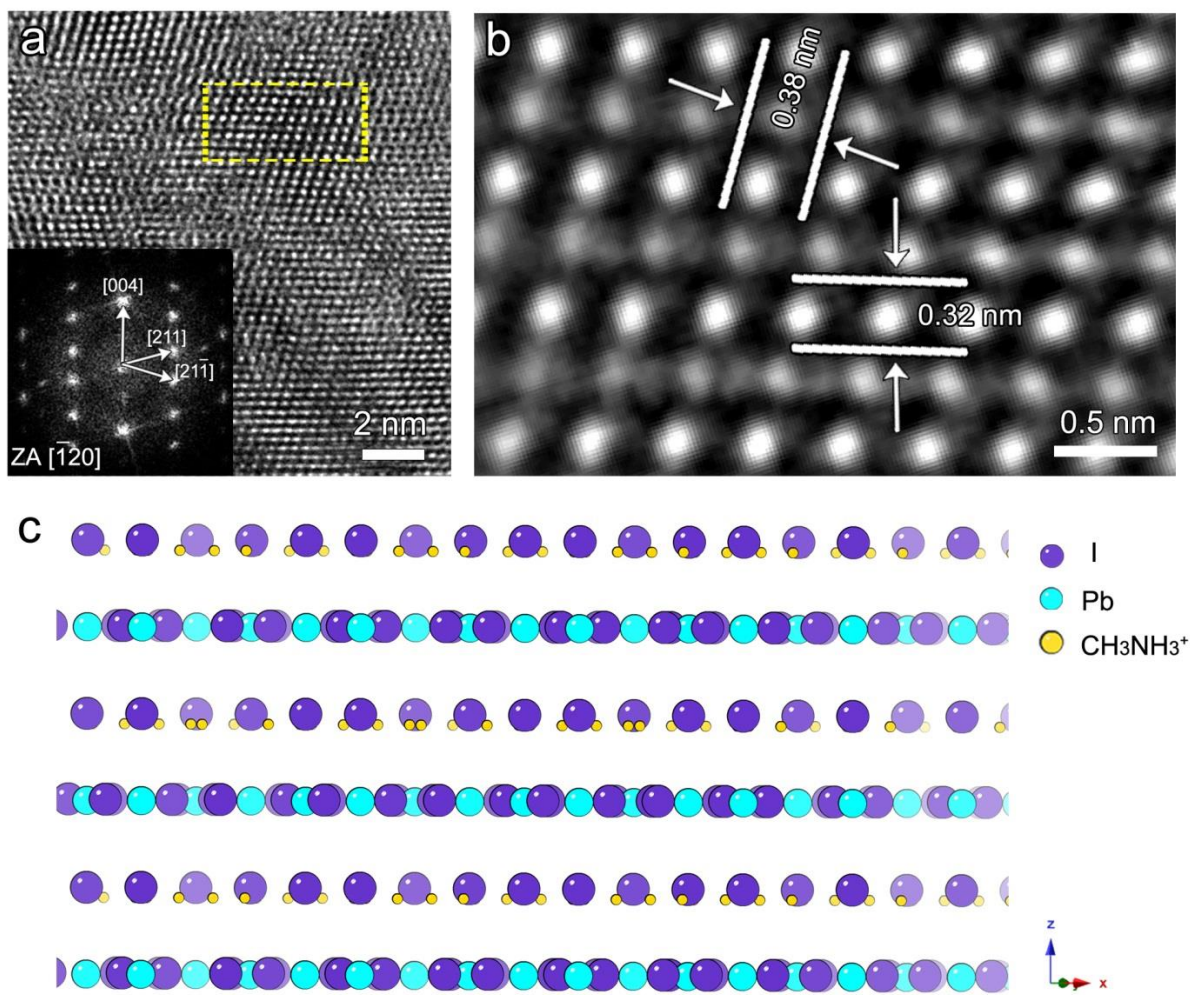

**Figure S13. High resolution TEM measurement on perovskite crystal nanoplatelets.** (a) High resolution TEM (HRTEM) image showing the structure of the perovskite nanoplatelets. Middle inset is the corresponding fast Fourier transform pattern from this HRTEM image along the [-120] zone axis (ZA). (b) HRTEM image of the selected-area marked by yellow in (a), with HRTEM filter process. (c) Schematic atomic structures of the perovskite crystal. Along x axis, there are alternative lines: one is consisting of I and Pb atoms while another consisting of I and $CH_3NH_3$ ions.

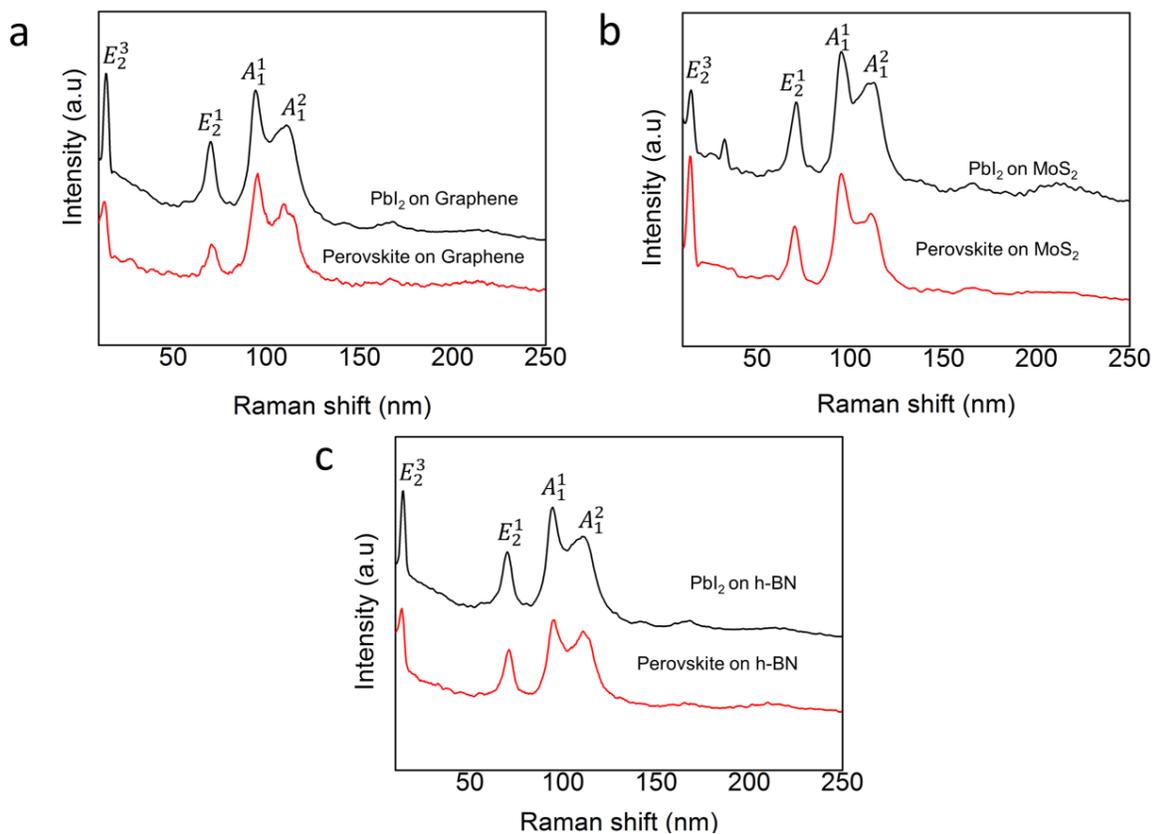

**Figure S14. Optical microscopy images of perovskite/2D vdW solids and corresponding Raman spectra of samples before and after conversion.** (d), (e) and (f) are Raman spectra of PbI$_2$ and perovskite nanoplatelets collected from the samples demonstrated in (a) (on graphene), (b) (on MoS$_2$) and (c) (on h-BN). In both PbI$_2$ and perovskite nanoplatelets, Raman spectra have the peaks at 14 cm$^{-1}$ assigned to $E_2^3$, at 70 cm$^{-1}$ assigned to $E_2^1$, at 94 cm$^{-1}$ assigned to $A_1^1$ and at 110 cm$^{-1}$ assigned to $A_1^2$.

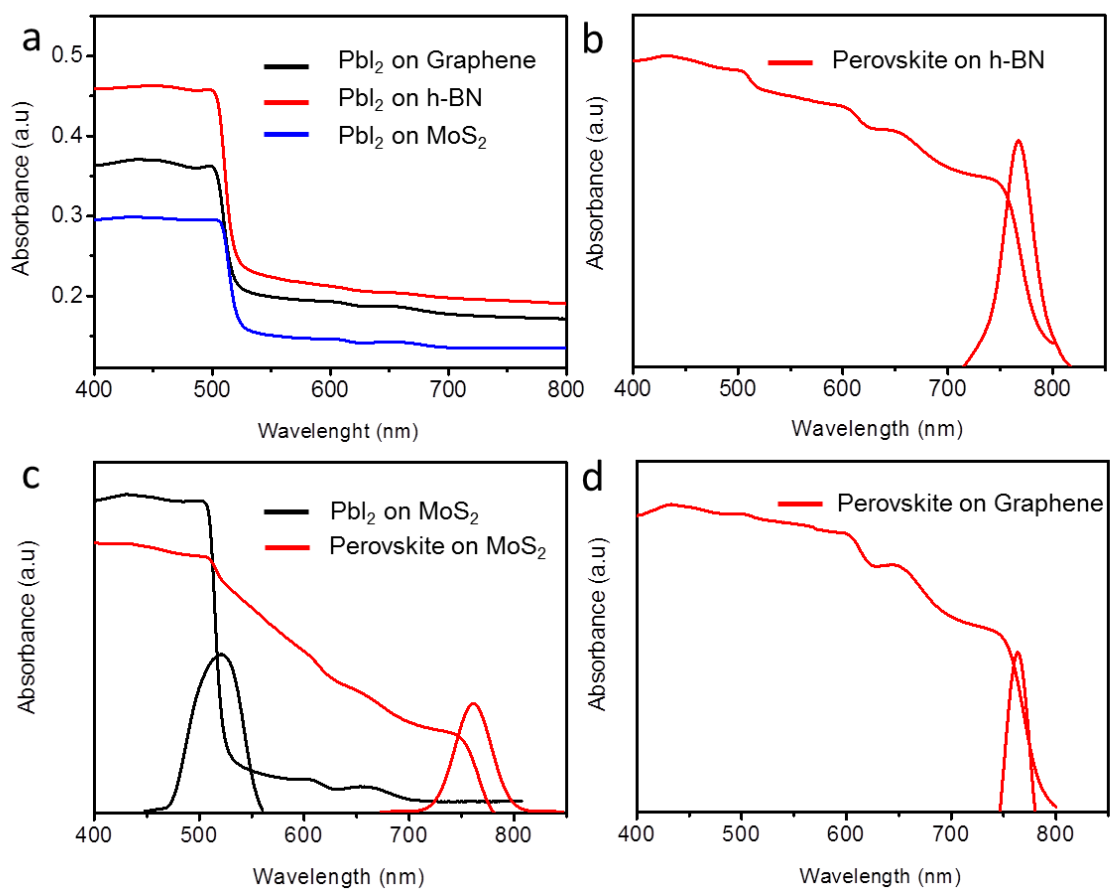

**Figure S15. UV/Vis absorption and PL excitation spectra of the same PbI₂ （before conversion） and perovskite/2D vdW (after conversion) solids grown on (a) graphene, (b) MoS₂ and (c) h-BN.** The black and red curves are optical spectra of PbI₂ and perovskite/2D vdW solids respectively. Before conversion, absorption peaks for PbI₂ nanoplatelets fall at around 500 nm whereas those of perovskite/2D vdW solids are around 760 nm. The positions of the excitation spectral features are in good agreement with those of the absorption spectra.

**S16-20 Discussion on the effect of three substrates on PbI$_2$ nanoplatelets.**

Being a novel class of vdW solids, it is important to understand the interface of this heterostructure, especially, the band structure and growing dynamics of PbI$_2$ on various 2D materials. In order to extract their band structure, we investigate the different possible positions of the PbI$_2$ relative to the 2D materials and their energies, by using first-principles calculations based on density functional theory (DFT) (Figures S17-19). We have examined energetics for multiple positions of the PbI$_2$ monolayer relative to the graphene, MoS$_2$ and *h*-BN. Supercells are used to reduce the lattice mismatch between PbI$_2$ and the 2D materials. Specifically, unit cell of PbI$_2$ on $\sqrt{3}\times\sqrt{3}$ supercell of graphene, 2×2 supercell of PbI$_2$ on 3×3 supercell of MoS$_2$, and unit cell of PbI$_2$ on $\sqrt{3}\times\sqrt{3}$ supercell of *h*-BN are adopted with lattice mismatch of 5.8%, 2.5% and 4.2%, respectively. For the case of graphene substrate (Figure S20a), hollow site configuration exhibits the lowest energy out of the 3 different possible positions. A further analysis indicates the interaction between monolayer PbI$_2$ and graphene is very weak, as shown in Figures S20b and S20c, that the band structure of PbI$_2$ does not display significant changes regardless of the presence of substrate. This is shown by the band diagram of PbI$_2$ on graphene (Figure S20b, in red) having the same profile as it is in its stretched standing alone case (Figure S20c, in red). The blue color plot in Figure S20c denotes the band diagram for a unit cell PbI$_2$ without strain effect used here as a reference. Moreover, this trend is also observed in systems of MoS$_2$ (seen Figures S20e and S20f) and *h*-BN (seen Figures S20h and S20c) substrates, which does not affect the pristine PbI$_2$ band structures in the corresponding supercells. In addition, it is realized that except for the slight changes in the energies of the systems, different positions of PbI$_2$ on the three 2D materials substrates does not considerably disturb the band structures of the overall systems (The results are presented in Figures S16-S20.)

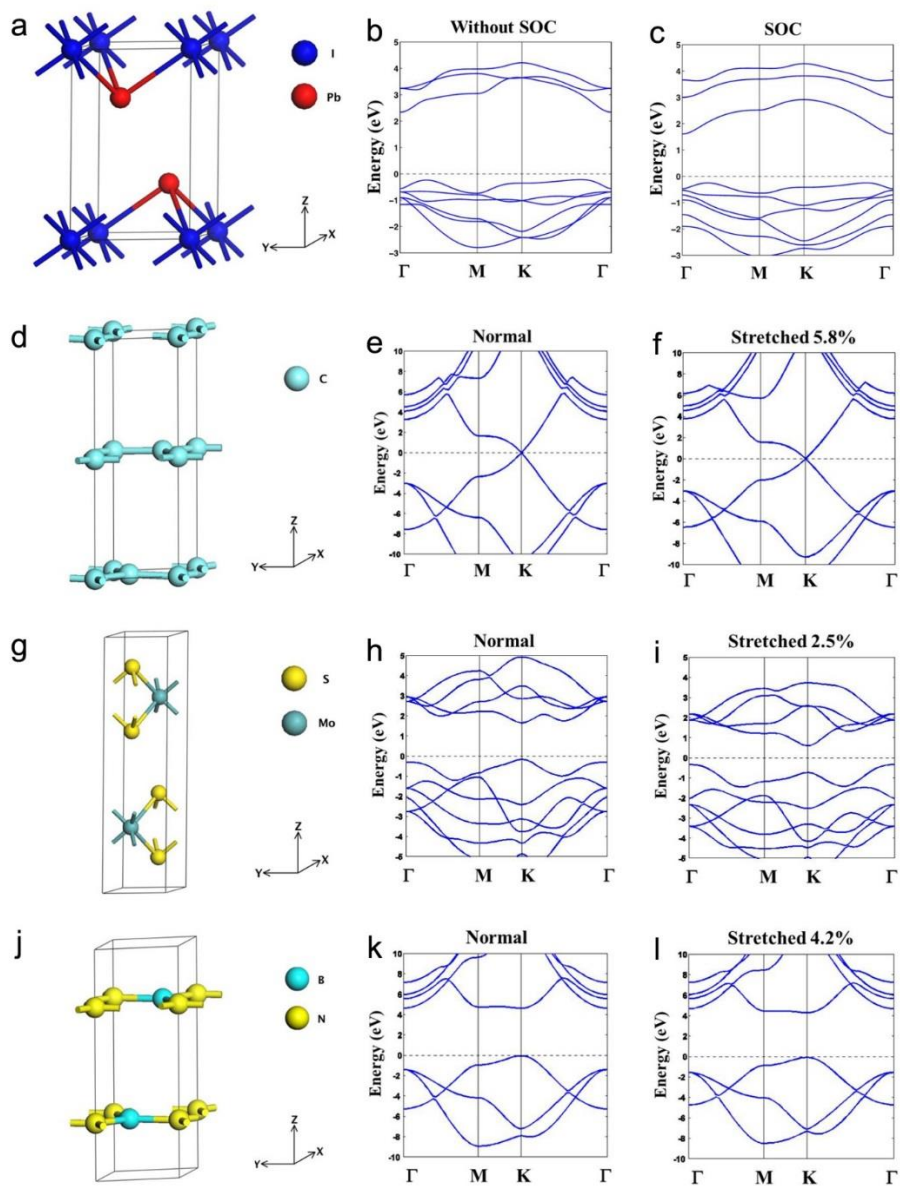

**Figure S16. Single crystal unit cell schematic and band structure of PbI2 and 2D materials.** Unit cell schematic: (a) PbI$_2$, (d) Graphene, (g) MoS$_2$, and (j) h-BN. With original crystal structures, the surface band structure of PbI$_2$ without SOC and the normal surface band structure of 2D materials: (b) PbI$_2$, (e) Graphene, (h) MoS$_2$, and (k) h-BN. To further understand the growth mechanism of perovskite/2D vdW solids, the calculation of lattice mismatch for epitaxial film is necessary. In view of lattice mismatch, the surface band structures of 2D materials with varying degrees of stretch: (f) Graphene with 5.8% stretch, (i) MoS$_2$ with 2.5% stretch, and (l) h-BN with 4.2% stretch. Comparing the band diagrams with and without stretching, the band profiles do not indicate significant changes when the unit cells are stretched. Hence,

theoretically a small strain on the substrates will not affect their band structures to a large extent. In addition, the effect of SOC on $PbI_2$ is only concentrates on the bottom few conduction bands, but not so much on the top few valence bands.

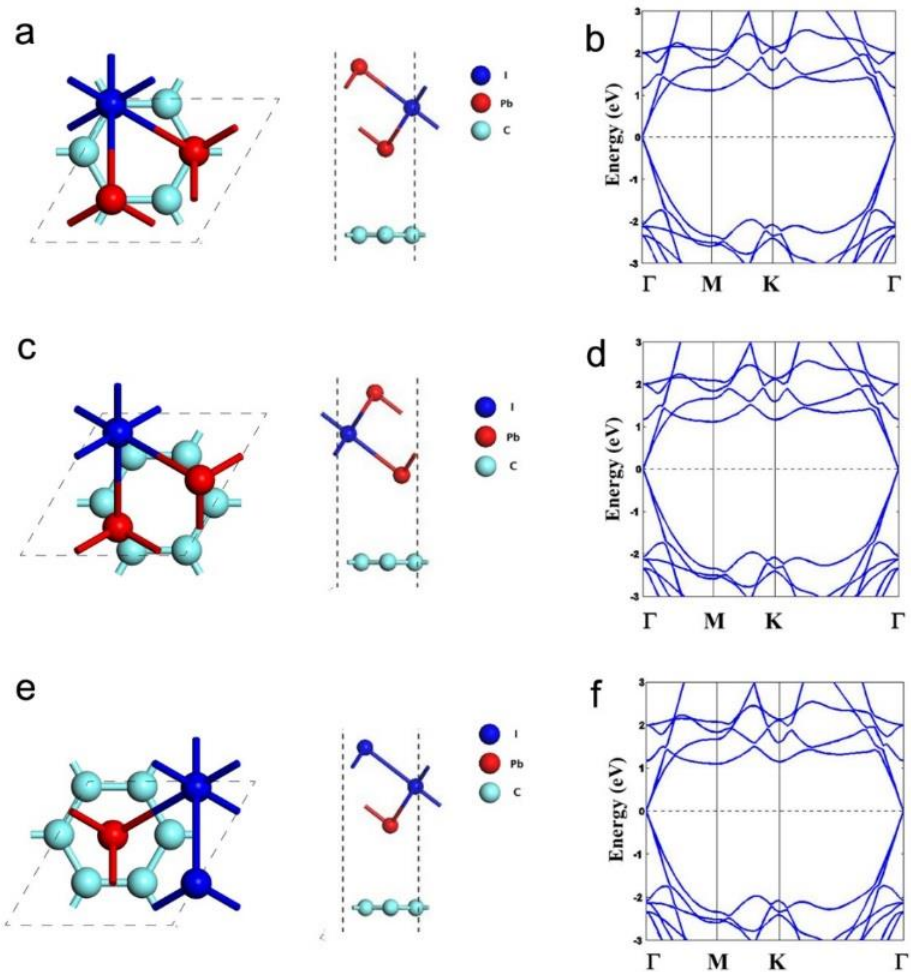

**Figure S17. Three possible positions and their corresponding surface band structures of the $PbI_2$ monolayer relative to the graphene substrate**: (a-b) Pb on C site, (c-d) bridge site, (e-f) hollow site. From the similar band profiles of the combined structures for the three configurations, it is clear that the system does not prefer one particular site to the others as they all have comparable energies. It is also noted that these combined band structures are merely a superposition of the band structures of $PbI_2$ and graphene standing alone without much distortion, indicating a weak interaction between the two layers.

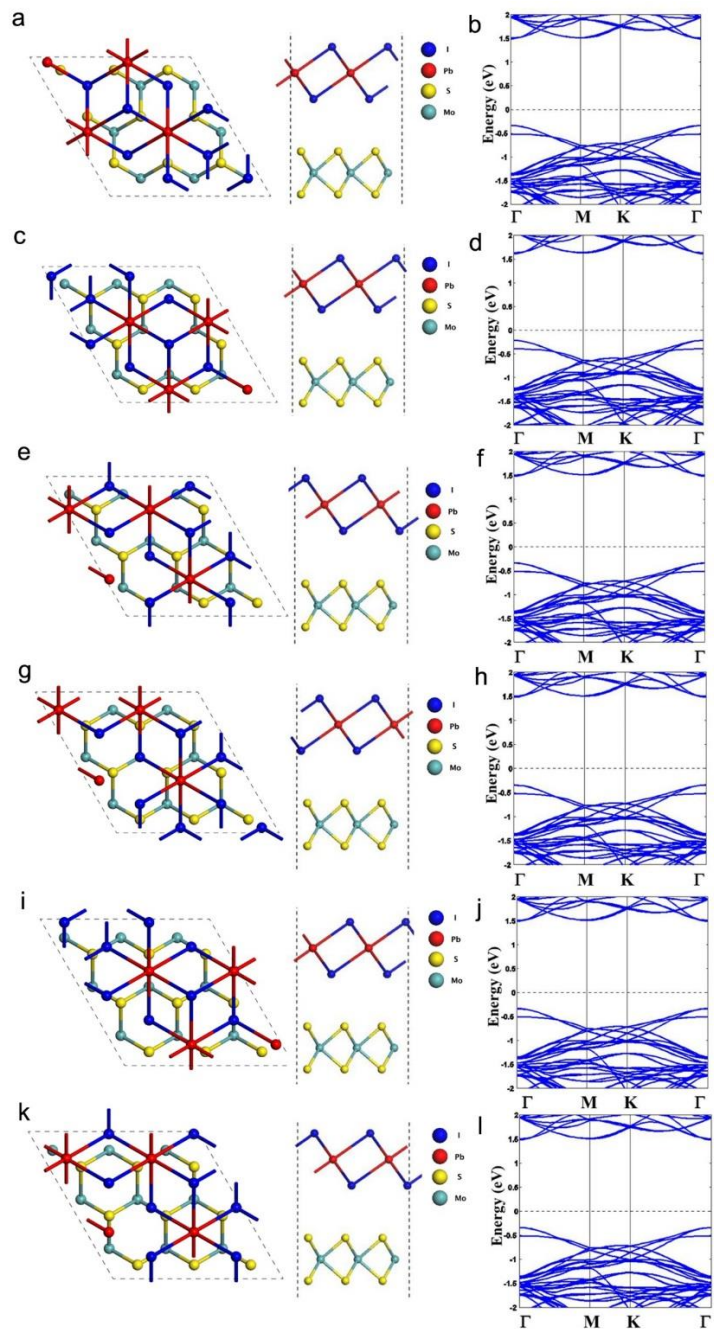

**Figure S18. Six possible positions and their corresponding surface band structures of the PbI$_2$ monolayer relative to the MoS$_2$ substrate:** (a-b) Pb on Mo site, (c-d) I on S site, (e-f) Pb on S site, (g-h) I on Mo site, (i-j) bridge site (Mo-Pb-S), and (k-l) hollow site. From the similar band profiles of the combined structures for the six configurations, it is clear that the system does not prefer one particular site to the others as they all have comparable energies as well as band gaps. It is also noted that these combined band structures are merely a superposition of the band structures of PbI$_2$ and MoS$_2$ standing alone without much distortion, indicating a weak interaction between the two layers.

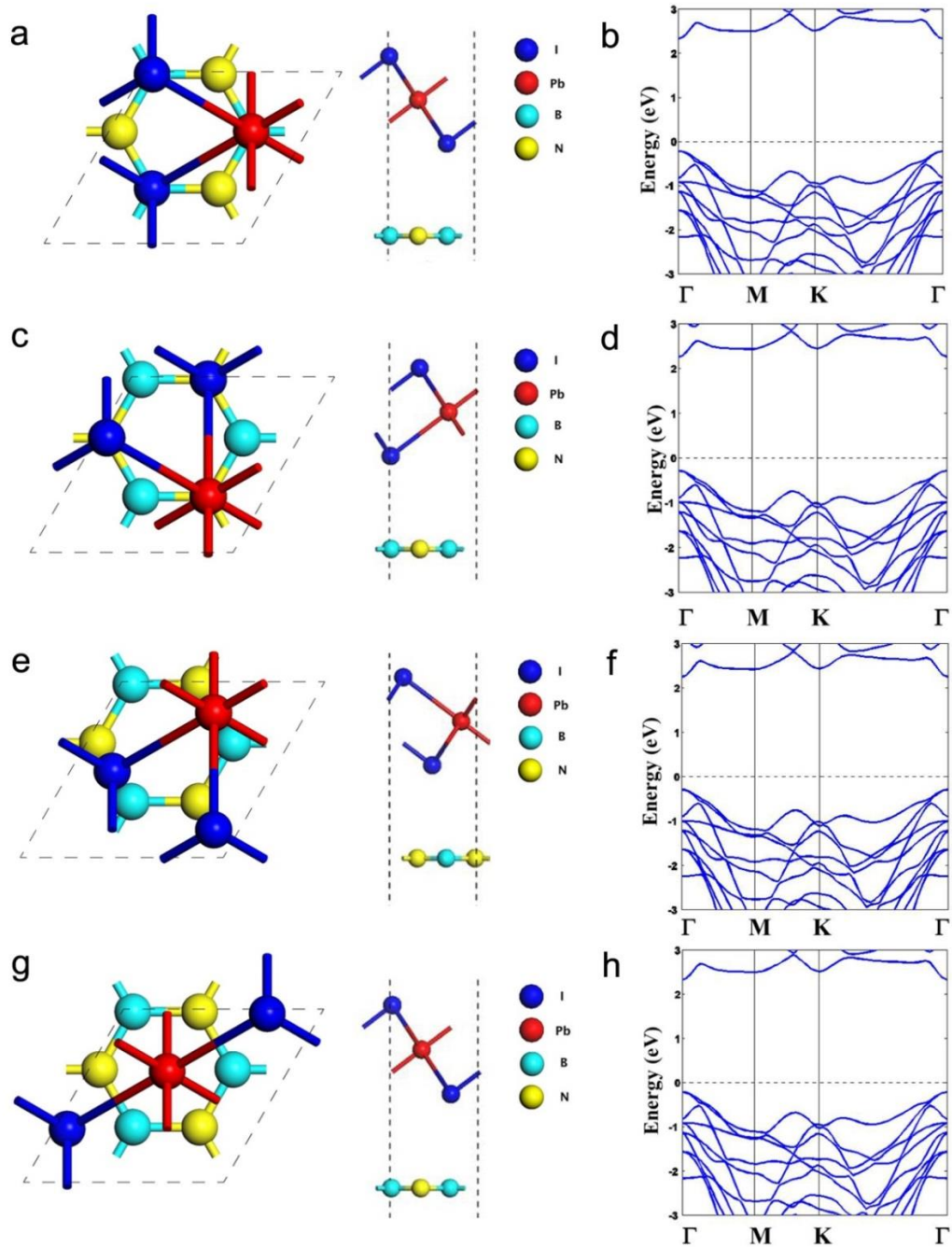

**Figure S19. Four possible positions and their corresponding surface band structures of the PbI$_2$ monolayer relative to the h-BN substrate:** (a-b) Pb on B site, (c-d) Pb on N site, (e-f) bridge site (B-Pb-N), and (g-h) hollow site. From the similar band profiles of the combined structures for the four configurations, it is clear that the system does not prefer one particular site to the others as they all have comparable energies as well as band gaps. It is also noted that these combined band structures are merely a superposition of the band structures of PbI$_2$ and h-BN standing alone without much distortion, indicating a weak interaction between the two layers.

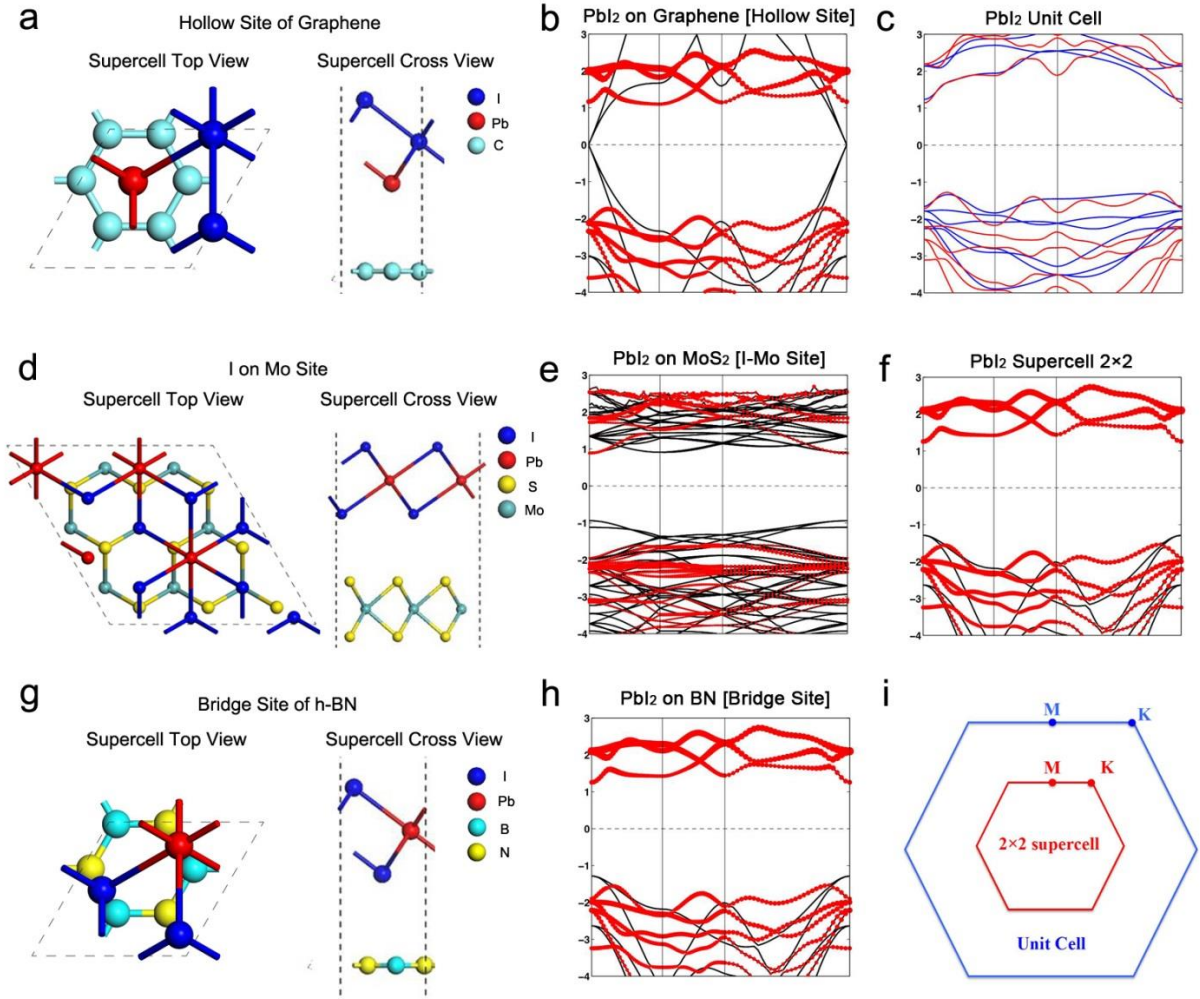

**Figure S20. Effect of three substrates on monolayer PbI$_2$.** (a) (d) (g) Lowest energy configurations of PbI$_2$ on various substrates: (a) unit cell of PbI$_2$ on √3 × √3 supercell of Graphene; (d) 2 × 2 supercell of PbI$_2$ on 3 × 3 supercell of MoS2; (g) unit cell of PbI$_2$ on √3 × √3 supercell of h-BN. (b) (e) (h) Band structures of the combined systems of monolayer PbI$_2$ and 2D substrates for the corresponding lowest energy sites: (b) Graphene for hollow site; (e) MoS$_2$ for I-Mo site; (h) h-BN for bridge site. The size of the red dots is proportional to the spectral weight from PbI$_2$ layer; (c) and (f) Band structures of 1 × 1 unit cell and of 2 × 2 supercell pristine monolayer PbI$_2$, respectively. In (c), the blue colored line represents

band profile for a normal unit cell of PbI$_2$ and the red colored line represents band profile for a stretched unit cell of PbI$_2$. (i) 1 × 1 and 2 × 2 supercells of hexagonal Brillouin zones.

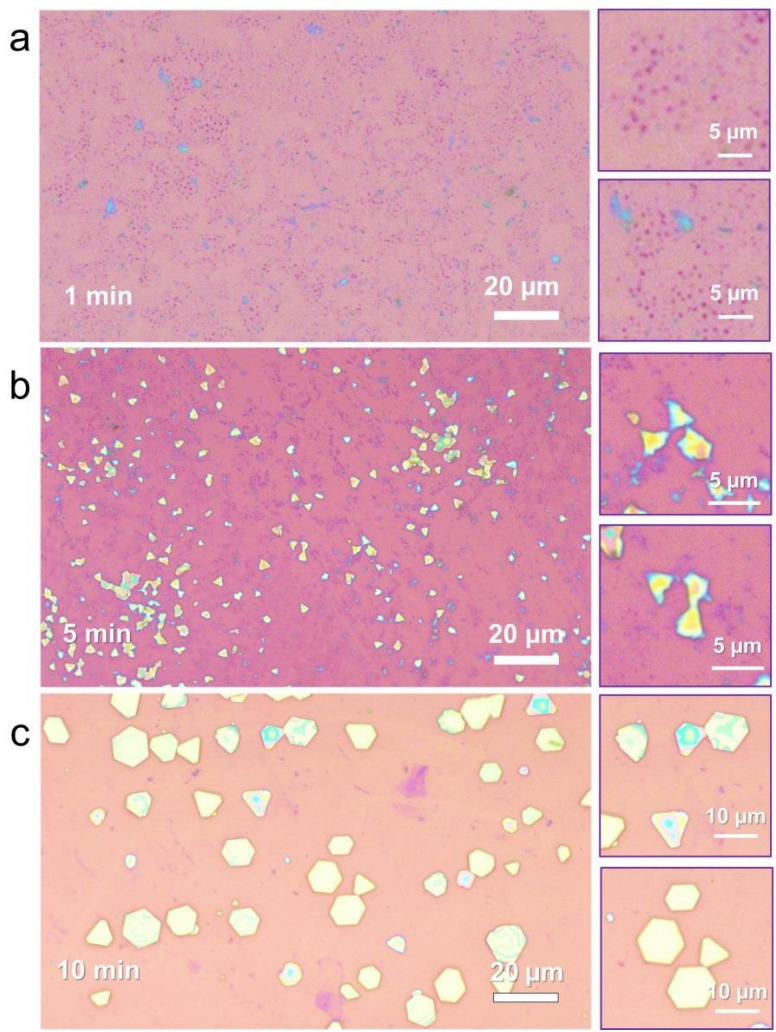

**Figure S21. Time-dependent growth of perovskite/2D vdW solids on CVD h-BN monolayer.** The samples with different reaction time ($t_R$) were synthesized and characterized by optical microscopy to investigate the evolution of PbI$_2$ nanoplatelets. (a) Small red dots observed at the early stage of the reaction ($t_R$= 1 min). (b) PbI$_2$ nanoplatelets with irregular shape formed and parts of these PbI$_2$ nanoplatelets tend to coalesce with each other ($t_R$= 5 min). (3) Hexagon shaped PbI$_2$ nanoplatelets formed ($t_R$= 10 min). On the basis of gradual morphology evolution, the growth of PbI$_2$ nanoplatelet was supposed to be formed through a van de Waals epitaxy mechanism. According to the time dependent experiments, there are two key points in the growth of perovskite/2D vdW solids: nucleation site and van de waals epitaxy.

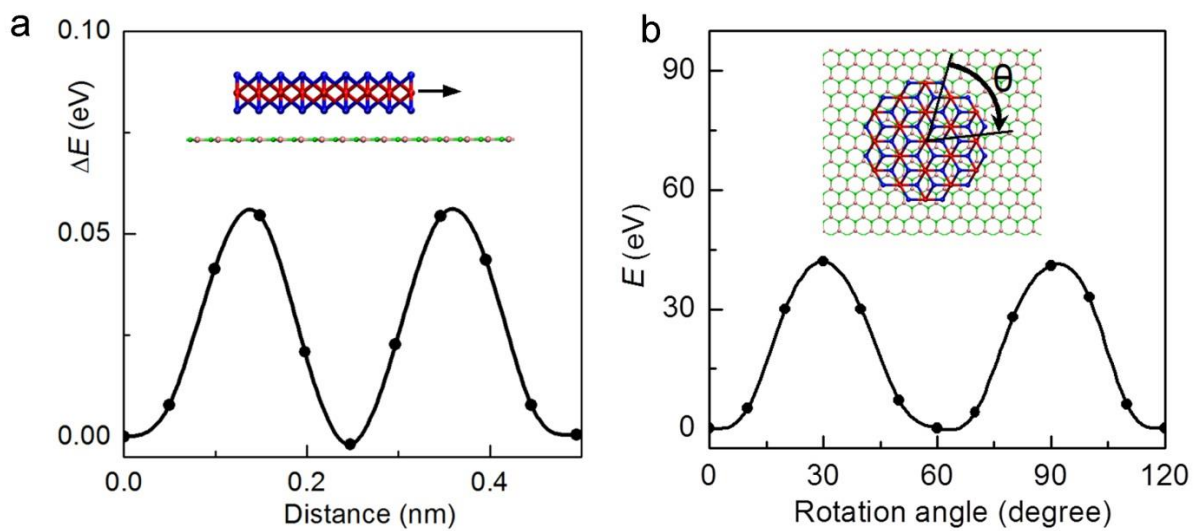

**Figure S22** (a) Potential energy as a function of sliding distance of a 2 nm-diameter nanoflake on a 2D *h*-BN sheet. (b) Potential energy as a function of rotation angle of a 2 nm-diameter nanoflake on a 2D *h*-BN sheet. The energies are calculated by first-principles calculations as implemented in VASP code. We employed ultrasoft pseudo-potentials for the core region and spin-unpolarized density functional theory based on local density approximation. A kinetic energy cutoff of 400 eV is chosen for the plane-wave expansion.

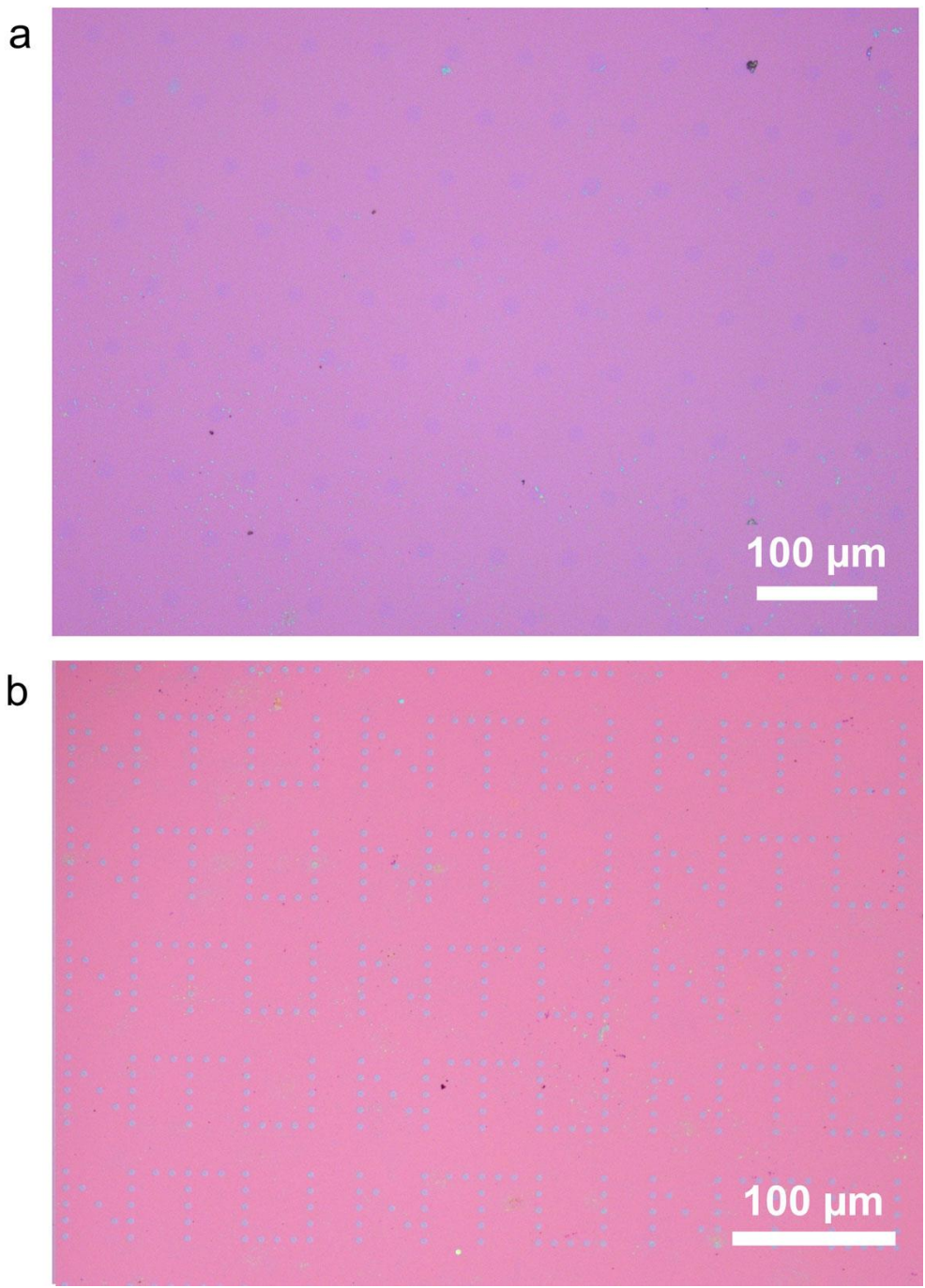

**Figure S23. Optical microscopy images of patterned h-BN.** (a) Highly ordered hexagonal arrays of h-NB patches formed on SiO$_2$/Si wafer. The diameter of hexagonal patch is 20μm and the gap distance between all adjacent patches is 50 μm. (b) NTU pattern of h-BN prepared on SiO$_2$/Si. The diameter of h-BN hexagonal patches in the NTU pattern is 10μm and the gap distance between all adjacent patches is 10 μm. The gap distance between all adjacent NTU units is 30 μm.

**Table S1 Assignments for low frequency Raman peaks of Perovskite/2D vdW solids**

|  | Frequency (cm$^{-1}$) | Symmetry |
|---|---|---|
| Perovskite/2D vdW solids | 14 | $E_2^3$ |
|  | 70 | $E_2^1$ |
|  | 94 | $A_1^1$ |
|  | 110 | $A_1^2$ |